\documentclass[prd,twocolumn,nofootinbib]{revtex4}
\usepackage{graphicx, epsfig}
\usepackage{color}
\usepackage{mathrsfs}
\usepackage{bm}
\usepackage{amsmath}
\usepackage[stable]{footmisc}

\newcommand{\be}{\begin{equation}}
\newcommand{\ee}{\end{equation}}
\newcommand{\ba}{\begin{eqnarray}}
\newcommand{\ea}{\end{eqnarray}}
\newcommand{\nn}{\nonumber}
\newcommand{\gapp}{\mathrel{\raise.3ex\hbox{$>$}\mkern-14mu
              \lower0.6ex\hbox{$\sim$}}}
\newcommand{\lapp}{\mathrel{\raise.3ex\hbox{$<$}\mkern-14mu
              \lower0.6ex\hbox{$\sim$}}}

\newcommand{\itie}{{\it i.e.}}
\newcommand{\bfA}{{\bf A}}
\newcommand{\bfB}{{\bf B}}
\newcommand{\bfBc}{{\bf B}_c}
\newcommand{\bfE}{{\bf E}}
\newcommand{\bfEc}{{\bf E}_c}
\newcommand{\bfJ}{{\bf J}}
\newcommand{\bfS}{{\bf S}}
\newcommand{\bfa}{{\bf a}}
\newcommand{\bfb}{{\bf b}}

\newcommand{\bfk}{{\bf k}}
\newcommand{\bfl}{{\bf l}}

\newcommand{\rms}{{\rm s}}

\newcommand{\bfv}{{\bf v}}
\newcommand{\bfr}{{\bf r}}
\newcommand{\bfx}{{\bf x}}
\newcommand{\rmG}{{\rm G}}
\newcommand{\muG}{\mu{\rm G}}
\newcommand{\kin}{k_{\rm in}}
\newcommand{\iteg}{{\it e.g.}}
\newcommand{\pc}{{\rm pc}}
\newcommand{\kpc}{{\rm kpc}}
\newcommand{\Gpc}{{\rm Gpc}}
\newcommand{\Mpc}{{\rm Mpc}}
\newcommand{\TeV}{{\rm TeV}}
\newcommand{\GeV}{{\rm GeV}}
\newcommand{\MeV}{{\rm MeV}}
\newcommand{\eV}{{\rm eV}}
\newcommand{\keV}{{\rm keV}}
\newcommand{\cm}{{\rm cm}}

\newcommand{\eepair}{$e^+e^-$}
\newcommand{\la}{\langle}
\newcommand{\ra}{\rangle}

\definecolor{bittersweet}{rgb}{1.0, 0.44, 0.37}
\definecolor{coolblack}{rgb}{0.0, 0.18, 0.39}
\definecolor{britishracinggreen}{rgb}{0.0, 0.26, 0.15}
\definecolor{coolgrey}{rgb}{0.55, 0.57, 0.67}
\definecolor{darkgreen}{rgb}{0.0, 0.2, 0.13}
\definecolor{darkmagenta}{rgb}{0.55, 0.0, 0.55}
\definecolor{eggplant}{rgb}{0.38, 0.25, 0.32}
\definecolor{fashionfuchsia}{rgb}{0.96, 0.0, 0.63}

\begin{document}
\title{Progress on Cosmological Magnetic Fields}
\author{Tanmay Vachaspati}
\affiliation{
Physics Department, Arizona State University, Tempe, AZ 85287, USA.
}

\begin{abstract}
\noindent
A variety of observations impose upper
limits at the nano Gauss level on magnetic fields that are coherent on inter-galactic scales
while blazar observations indicate a lower bound $\sim 10^{-16}$ Gauss.
Such magnetic fields can play an important astrophysical role, for example at cosmic recombination 
and during structure
formation, and also provide crucial information for particle physics in the early universe. Magnetic 
fields with significant energy density could have been produced at the electroweak phase transition. 
The evolution and survival of magnetic fields produced on 
sub-horizon scales in the early universe, however, depends on the magnetic helicity which is related 
to violation of symmetries in fundamental particle interactions. The generation of 
magnetic helicity requires new CP violating interactions that can be tested by 
accelerator experiments via decay channels of the Higgs particle. 
\end{abstract}

\maketitle

\section{Introduction}
\label{introduction}

Several excellent reviews on cosmological magnetic fields
exist~\cite{Grasso:2000wj,Widrow:2002ud,Vallee:2004osq,Durrer:2013pga,Subramanian:2015lua}.
This article is a perspective on where the subject is in 2020, on several claims and
counter-claims, and on open problems of interest for the future. The focus is on
physical aspects of the generation of cosmological magnetic fields, their evolution, 
and observation.

Let us start by discussing the generation of magnetic fields.
The presence of a cosmological plasma suggests that the discussion 
should be based in the language of magneto-hydrodynamics (MHD), with
the MHD equation ({\it e.g.} Ch.~10 in~\cite{Jackson})\footnote{We use
Maxwell equations in Lorentz-Heaviside units in this article.},
\be
\partial_t \bfB = \nabla \times (\bfv \times \bfB ) + \frac{1}{\sigma}\nabla^2\bfB
\ee
where $\bfB$ is the magnetic field, $\bfv$ the plasma velocity, and $\sigma$ the 
electrical conductivity of the plasma. The key point is that the plasma does not provide
a source term 
for the magnetic field: if $\bfB (t=0,\bfx)=0$ then $\bfB(t,\bfx)=0$ for all times and
magnetic fields cannot be generated within the MHD description.

Fortunately source terms can be present when we go beyond MHD.
The charges in standard astrophysical plasmas consists of electrons and protons, which have
equal and opposite electric charge but the masses are vastly different, $m_p \approx 2000 m_e$. 
The Thomson interaction cross-sections are also vastly different: 
$\sigma_{e\gamma}/\sigma_{p\gamma} \approx (m_p/m_e)^2 \sim 10^6$.
This opens up the possibility for generating electric currents when astrophysical plasmas interact with
photons. For example, consider rotational flow of the cosmological medium prior to recombination.
 The electrons scatter more efficiently with the ambient radiation and feel a greater drag than do the 
 protons, producing a net electric current that sources a magnetic field. 
Harrison~\cite{1970MNRAS.147..279H,Pogosian:2001np} has used this scheme to propose 
(weak) magnetic field generation if there is turbulence during cosmological 
recombination\footnote{In astrophysical scenarios, a source term in the MHD equation is provided 
by the ``Biermann battery'' term which is proportional to $\nabla n_e \times \nabla p$ where $n_e$
is the electron number density and $p$ is the fluid pressure. In certain astrophysical situations
this term can become non-zero. It has also been applied during the cosmological QCD epoch
in~\cite{Quashnock:1988vs}.}.

The Harrison mechanism illustrates how the violation of fundamental symmetries might play
a role in the generation of magnetic fields. Crucial use is made of the violation of electric charge 
conjugation symmetry (C) since electrons and protons have different masses and interaction 
strengths with photons\footnote{The connection between symmetry 
violations and magnetogenesis was first considered in analogy with the Sakharov conditions 
for baryogenesis in Ref.~\cite{Davidson:1996rw}.}. 
The role of fundamental symmetry violations becomes even more interesting in the context of 
early universe cosmology and the electroweak phase transition serves as an illustration. At
the electroweak epoch the cosmological medium has essentially equal numbers of particles 
and antiparticles
and there is only very weak CP violation that distinguishes their interactions.
Instead, as discussed in Sec.~\ref{production}, the production of magnetic fields follows from
the dynamics of the Higgs field during the phase transition and it appears that the violation
of fundamental symmetries is not necessary. 
However, there is more to the story, as the evolution and survival of cosmological magnetic 
fields depends critically on the helicity (circular polarization) of the magnetic field. 
A helical magnetic field is characterized by the quantity $\bfB\cdot \nabla\times \bfB$
and overall magnetic helicity implies that the average value of $\bfB\cdot \nabla\times \bfB$
is non-vanishing, in turn implying a violation of parity (P) and charge+parity (CP) symmetries. 
Therefore the present amplitude of cosmological magnetic fields that were generated 
at the electroweak epoch depends on the strength of P and CP symmetry violations
in the fundamental interactions, just as the amount of cosmic matter-antimatter asymmetry 
depends on violations of these symmetries.
Then the observation of cosmological magnetic fields
and their helicity can inform us about fundamental symmetry violations that should also 
appear in accelerator experiments, thus making another important
outer-space/inner-space connection.

In the cosmological context, Maxwell electrodynamics is only applicable after the 
electroweak phase transition (EWPT), $\sim 10^{-10}\,\rms$ after the big 
bang\footnote{I refer to the spontaneous breaking of electroweak symmetry
as a ``phase transition'', disregarding the dynamics, whether it is a first or second
order transition, or a smooth cross-over as in the standard model.}. 
Prior to the
EWPT, there were 3 SU(2) (``weak'') gauge fields and 1 U(1) (``hypercharge'') gauge field
and electromagnetism was undefined. 
Once the electroweak $SU(2)\times U(1$) symmetry is broken, only one of the gauge fields 
remains massless
and is what we call the electromagnetic gauge field. Although the electroweak model may be
unfamiliar to non-particle physicists, since the discovery of the Higgs particle the model is about 
as robust as Maxwell's electrodynamics. There are sure to be some extensions but the basic 
equations are firm. So predictions from the electroweak equations can be made with 
similar confidence as from Maxwell equations, and if the electroweak equations generate
magnetic fields at the EWPT, they are as real as say the cosmic microwave background.

I will discuss the generation of magnetic fields at the EWPT in some detail in~Sec.~\ref{ewptB}.
It is important to note that electromagnetism as derived from the electroweak model has
magnetic sources (magnetic monopoles) and ${\rm div}(\bfB) \ne 0$ in general. Once
the Higgs has acquired a non-zero vacuum expectation value (VEV) everywhere and 
the EWPT is complete, we recover Maxwell's 
equations and MHD. During the EWPT the full electroweak dynamics needs to be taken into 
account. This opens up new research problems -- How can one include plasma effects in
the electroweak model? Is there a suitable ``non-Abelian MHD'' approximation applicable to
the electroweak plasma? How do results depend on ``unknowns'' such as neutrino masses, 
dark matter, additional sources of CP violation? 

The ultimate test of all theoretical ideas is whether they are confirmed by experiment
or observations. For cosmological magnetic fields there has been a lot of progress 
on the observational front. Some decades ago, Faraday rotation of quasars placed
the most stringent constraints on cosmological magnetic fields. Now there are additional
constraints from the cosmic microwave background and from blazar gamma ray data.
Importantly, blazar observations place {\it lower} bounds on magnetic fields provided
our understanding of electron/positron propagation in the cosmological medium is
correct. Blazar observations can in principle be used to detect magnetic helicity and
the spectra of magnetic fields. This topic is discussed in Sec.~\ref{observations}.

The presence of magnetic fields may impact our understanding of other
cosmological epochs. Already there is discussion of whether magnetic fields can
affect faster Hydrogen recombination and potentially resolve the tension in the
low redshift Hubble constant measurements with those using the 
CMB~\cite{Jedamzik:2020krr}.
The interactions of hypothesized neutrino magnetic moments with primordial magnetic
fields may affect big bang nucleosynthesis~\cite{Enqvist:1993sa,Enqvist:1994mb} and
have consequences for neutrino detection experiments~\cite{baym2021evolution}.
There may be other effects waiting to be discovered -- do magnetic fields affect the
QCD phase transition? do they leave an imprint on the 21 cm observations?
do they affect structure formation? And, very importantly, can they help explain 
the observed magnetic fields in galaxies and clusters of galaxies?

There already are attempts to answer a large number of these questions. As in
all good science, there are claims and counter-claims, making this a fertile
ground for research, innovative ideas, and further experiments and observations.
The purpose of this progress report is to give a perspective on the salient results
and discussions. 

After setting up some basic conventions and notation in Sec.~\ref{framework},
I start in Sec.~\ref{observations} by discussing some (not all)
observational efforts, focusing on blazar observations as these appear to be
most promising at present. In Sec.~\ref{production}, I discuss the generation
of magnetic fields in the early universe, focusing on the production during
the EWPT. Then I come to a summary of the evolution of cosmological magnetic
fields in Sec.~\ref{evolution}. Finally, in Sec.~\ref{ideasforamplifying}, I discuss a few
other ideas for generating and amplifying magnetic fields. Here I also discuss
the possibility that magnetic fields may indicate new fundamental interactions
and that these could be tested in accelerator 
experiments\footnote{I will not be discussing the scenario where magnetic fields 
are assumed as an initial condition leading to consequences for 
baryogenesis~\cite{Dvornikov:2011ey,Dvornikov:2012rk,Fujita:2016igl,
Kamada:2016cnb,Kamada_2016}.}.

We use natural units ($\hbar=1=c$) throughout, Lorentz-Heaviside conventions 
for electromagnetism, and a flat  Friedman-Robertson-Walker cosmology.
For numerical estimates, 
conversions between different units are neatly summarized in~\cite{conversion}.

\subsection{Framework: stochastic, statistically isotropic magnetic fields}
\label{framework}

We can imagine a uniform magnetic field that pervades the universe. However, this
cannot be the outcome of a local, dynamical process since locality implies that 
magnetic field directions in vastly separated regions are governed by independent
physics. We will only consider such local processes in this report. Then we are
interested in stochastic magnetic fields that on average are isotropic. The
correlation function for any divergence-free stochastic vector field that is statistically 
isotropic can be written as \cite{MoninYaglom},
\be
\la B_i (\bfx +\bfr) B_j (\bfx ) \ra =
M_N(r) P_{ij} + M_L(r) {\hat r}_i{\hat r}_j + \epsilon_{ijk} r_k M_H(r)
\label{BBcorr}
\ee
where $P_{ij}=\delta_{ij}-{\hat r}_i {\hat r}_j$ is the projection tensor orthogonal
to ${\hat r}$; $M_N$, $M_L$ and $M_H$ are respectively the ``normal'', 
``longitudinal'' and ``helical'' parts of the correlation function\footnote{Attention 
should be paid to the order of the points on the
left-hand side of \eqref{BBcorr} and the signs and factors on the right-hand
side since different conventions are prevalent in the literature.}.
Since $\nabla \cdot \bfB =0$, we have a relation between the
normal and longitudinal spectra
\be
M_N(r) = \frac{1}{2r} \frac{d}{dr} \left ( r^2 M_L(r) \right )
\label{MNML}
\ee
With the Fourier transform conventions\footnote{$\delta(x) = \int dk\, e^{-ikx}/(2\pi)$},
\be
\bfb(\bfk) = \int d^3 x\, \bfB (\bfx) e^{+i\bfk \cdot \bfx}, \ 
\bfB(\bfx) = \int \frac{d^3 k}{(2\pi)^3} \bfb (\bfk) e^{-i\bfk \cdot \bfx}
\label{ft}
\ee
the k-space correlator takes the form,
\ba
\la b_i (\bfk) b_j^* (\bfk ') \ra = 
\left [ \frac{E_M(k)}{4\pi k^2} p_{ij} + i \epsilon_{ijl}k^l \frac{H_M(k)}{8\pi k^2} \right ]
&&
\nn \\
&& \hskip -3 cm
\times (2\pi)^6 \delta^{(3)} (\bfk - \bfk' )
\label{bbcorr}
\ea
where $p_{ij}=\delta_{ij}-{\hat k}_i {\hat k}_j$. $E_M(k)$ is called the
``power spectrum'' and $H_M(k)$ the ``helicity spectrum'' (or sometimes
the ``helicity power spectrum''). Commonly encountered spectra in the
cosmology literature are the ``Batchelor spectrum'' with $E_M(k) \propto k^4$ 
and the ``scale invariant'' spectrum with $E_M(k) \propto k^{-1}$.

The correlators refer to an ensemble
average over many stochastic realizations of the magnetic field. To
connect theory and observations, since we only have one realization 
of the magnetic field in the universe, the ensemble average is to be 
thought of as a spatial average (the ``ergodic hypothesis''),
\be
\la B_i (\bfx +\bfr) B_j (\bfx ) \ra \to 
\frac{1}{V} \int_V d^3x \, B_i (\bfx +\bfr) B_j (\bfx ) 
\ee
where $V$ is some large volume. This assumes that the spatial
separations ($r$) of interest are smaller than the extent of the integration
volume.

The $x$-space and $k$-space correlation functions can be related,
\be
M_N(r) + \frac{1}{2} M_L (r)  =   \int_0^\infty dk \, E_M(k) \frac{\sin(kr)}{kr}
\label{MNMLEMk}
\ee
with $M_N$ and $M_L$ related as in \eqref{MNML}, and
\be
M_H(r) =  - \frac{1}{2r} \int_0^\infty dk \, k H_M(k) 
             \frac{d}{d(kr)} \left ( \frac{\sin (kr)}{kr} \right )
\label{MHHMk}
\ee

A clarification regarding ``magnetic helicity'' is necessary. Magnetic helicity
is defined as
\be
{\rm magnetic \ helicity} \equiv \int d^3x \, \bfA \cdot \bfB
\ee
where the integral is over all space. (In a finite volume, the expression is gauge 
invariant provided the magnetic field is orthogonal to the areal vector everywhere 
on the boundary of the volume.) Magnetic 
helicity has an interpretation in terms of the linking number (or writhe and twist) of 
magnetic flux lines~\cite{Moreau:1961,Moffatt:1969,BergerMoffatt:1984}.
Magnetic helicity is a very useful quantity because it is conserved in MHD
evolution of plasmas with high electrical conductivity. Even if the electrical 
conductivity is finite, helicity conservation is observed in numerical solutions
and has been used to explain experimental results~\cite{PhysRevLett.33.1139}. 
However, magnetic helicity is a non-local quantity and hence is not experimentally
measurable.  Instead the ``physical magnetic helicity'' defined as
$\bfB \cdot \nabla \times \bfB$ is local and measurable. From \eqref{BBcorr}
we derive,
\be
\la \bfB \cdot \nabla \times \bfB \ra = 6 M_H(0) = \int_0^\infty dk\, k^2 H_M(k)
\label{BcurlB}
\ee
where the last relation can be derived using \eqref{MHHMk}.

Observations only probe an averaged value of inter-galactic magnetic fields.
Since the magnetic field is a three-vector, there are several ways to define
the ``average magnetic field strength''~\cite{Enqvist:1993np,Hindmarsh:1997tj}.
For example, the line-averaged magnetic field strength on a curve of length $L$
may be defined as,
\be
{\bar B}_{\rm line} = \frac{1}{L} \int_0^L d\bfl \cdot \bfB
\ee
and one may further average over a set of such curves.
In the literature, one also encounters an ``effective'' magnetic field based 
on the averaged energy density,
\be
B_{\rm eff}^2 = \frac{1}{V} \int_V d^3x \,  \bfB^2
\label{Beff}
\ee
Following Durrer and Neronov~\cite{Durrer:2013pga}, we may also define the 
``magnetic field strength on scale $\lambda$'',
\be
B_\lambda \equiv \sqrt{2 k  E_M(k)}
\label{Blambda}
\ee
where $\lambda \equiv 2\pi/k$.
In contrast, the Planck collaboration~\cite{planck} defines the magnetic 
field strength smoothed on a scale $\lambda$ as
\be
{\cal B}_\lambda^2 = \int_0^\infty \frac{dk\, k^2}{2\pi^2} e^{-\lambda^2 k^2} P_B(k)
\label{calB}
\ee
where, to bridge conventions,
\be
P_B(k) = 2 (2\pi )^3 \frac{E_M(k)}{4\pi k^2}.
\ee
Another definition advocated in~\cite{Hindmarsh:1997tj} is the volume averaged magnetic
field,
\be
\bfB_V = \frac{1}{V} \int_V d^3x\, \bfB (\bfx ) .
\label{BV}
\ee
If the volume is chosen to be a sphere of radius $\lambda$ we can express the
root-mean-square of $\bfB_V$ in terms of $E_M(k)$,
\be
B_{V,\lambda}^2 = 2 \int dk \, E_M(k) W^2_V(k)
\label{BVlambda}
\ee
where
\be
W_V(k) = \frac{3}{(k\lambda )^3} (\sin(k\lambda ) - k \lambda \cos(k\lambda))
\ee
The window function $W^2_V(k)$ in \eqref{BVlambda} limits the integration to regions
with $k \lesssim 1/\lambda$, similar to the Gaussian window in \eqref{calB}.
For power law $E_M$ that are not too extreme, these different definitions of ``magnetic 
field strength on a scale $\lambda$'' agree with each other up to ${\cal O}(1)$ numerical 
factors.

Different observations are sensitive to differently averaged
magnetic fields. For example, Faraday rotation observations 
are sensitive to the line averaged magnetic field
${\bar B}_{\rm line}$ weighted by the electron number density (see Sec.~\ref{observations}), 
whereas big bang nucleosynthesis constraints 
depend on the average energy density as given by $B_{\rm eff}$. 
To derive constraints in a unified form, it seems best to recast all of these different
averages in terms of integrals over the power spectra convoluted with some
window function. However, this has not yet been done.

The average magnetic energy density can be written in terms of the
correlators,
\be
\rho_B = \frac{1}{2} \la \bfB^2 \ra = M_N(0) + \frac{M_L(0)}{2}
= \int dk \, E_M(k)
\label{Benergydensity}
\ee
Similarly we can define the average magnetic helicity density over a
large volume $V$. We first relate the Fourier transformed gauge field,
denoted $\bfa (\bfk)$, to $\bfb (\bfk )$,
\be
\bfa (\bfk) = -i \frac{\bfk}{k^2} \times \bfb(\bfk ) + \Lambda (\bfk ) \bfk
\ee
where $\Lambda (\bfk )$ is an arbitrary function that depends on the gauge
choice. Then, the average magnetic helicity is related to the helicity power
spectrum by,
\be
\la h \ra \equiv \frac{1}{V} \int_V d^3x \, \la \bfA \cdot \bfB \ra = \int dk \, H_M(k)
\label{hdefn}
\ee
where we have used $\bfk \cdot \bfb (\bfk)=0$ (equivalent to $\nabla\cdot \bfB=0$).

We have seen that stochastic, isotropic magnetic fields are described by
two power spectra: $E_M(k)$ related to the energy density, and $H_M(k)$
related to the helicity. However, there can't be any helicity without energy
and so the two spectra are weakly related~\cite{1978mfge.book.....M},
\be
E_M(k) \ge \frac{k}{2} | H_M(k) |
\label{realizability}
\ee
which can be derived using the Cauchy-Schwarz inequality. This condition
is called the ``realizability condition'' and magnetic fields that saturate this
condition are called ``maximally helical''. Later we will see that MHD evolution
tends to drive magnetic fields towards maximal helicity.

\subsection{Some numbers}
\label{somenumbers}

It is helpful to keep certain order-of-magnitude numerical values in 
mind. To start, the magnetic field at the surface of Earth is $\sim 1\, \rmG$, as is the 
magnetic field in the solar corona. Galactic magnetic fields are on the order of 
$10^{-6}\, \rmG$ and are coherent on kpc scales. Most constraints on cosmological 
magnetic fields give an upper bound around $10^{-9}\, \rmG$ assuming coherence 
on Mpc scales or larger. Claimed lower bounds on inter-galactic magnetic field strengths 
are on the order of $10^{-16}\, \rmG$ on Mpc scales.

In our discussion it will also help to keep in mind that the energy density in magnetic 
fields of $10^{-6}\, \rmG$ is $\sim (10^{-4}\, \eV)^4$ and is comparable to the energy density 
in photons at a temperature of $\sim 3\, {\rm K}$, which is also the temperature of the 
cosmic microwave background\footnote{We use $1\, \rmG = 1.95 \times 10^{-2}\, \eV^2$.}.
Thus the galactic magnetic field has energy density comparable to the cosmological 
radiation density.

\section{Observations}
\label{observations}

A variety of observational tools are employed to detect cosmological magnetic fields including 
Faraday Rotation (FR) of linearly polarized sources~\cite{Kronberg:1993vk,Blasi_1999,Pshirkov_2016}, 
deflection of cosmic rays~\cite{Lemoine:1997ei,Bertone:2002ks},
imprints on the temperature and polarization of the CMB~\cite{Kahniashvili:2000vm,Kosowsky:2004zh,
Campanelli:2004pm,
Kahniashvili:2008hx,Miyamoto:2013oua,Kahniashvili:2014dfa,planck,Hort_a_2014,Hort_a_2017,
Paoletti:2018uic,vazza2020simulations,Brandenburg_2018}, 
effects on light element abundances 
(BBN)~\cite{1970ApJ...160..451M,1969Natur.223..938G,Cheng:1993kz,
Grasso:1994ph,Kernan:1995bz,Cheng:1996yi,Kernan:1996ab}, and electromagnetic cascades 
from high energy blazars~\cite{Neronov73,Essey_2011,Finke_2015,Biteau:2018tmv,
korochkin2020sensitivity,AlvesBatista:2020oio}. 
In addition there are constraints on primordial magnetic
fields obtained from the structure of dwarf galaxies~\cite{sanati2020constraining}. 
Potentially, Fast Radio Bursts (FRB)~\cite{Hackstein:2019abb} and Gamma Ray Bursts (GRB) may 
also be used to probe inter-galactic magnetic
fields~\cite{Neronov73,2011MNRAS.414.3566T,Wang:2020vyu,Dzhatdoev:2020yvn}.
In the near future, we can expect more observational results from the Square Kilometer 
Array (SKA)~\cite{Heald:2020pgv} and the Cherenkov Telescope Array (CTA)~\cite{consortium2020sensitivity}.

\begin{figure}
      \includegraphics[width=0.44\textwidth,angle=0]{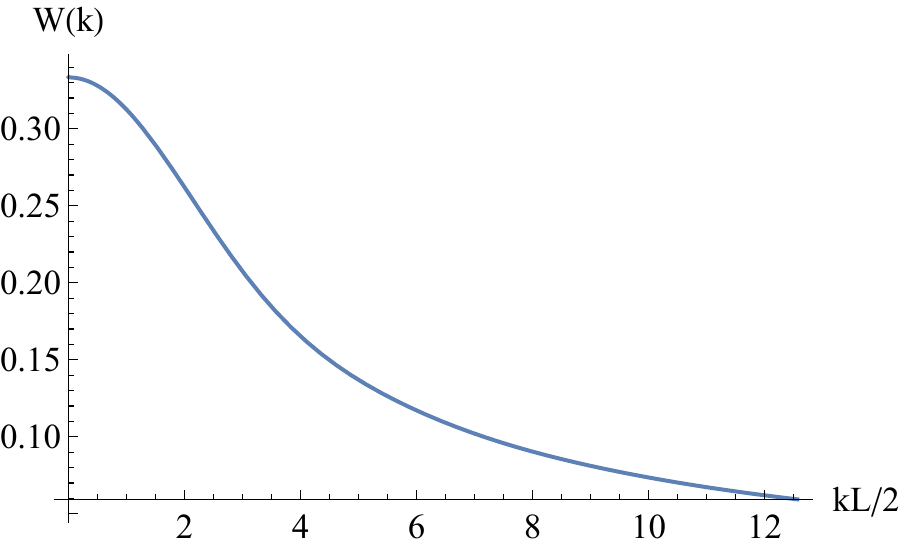}
        \caption{The window function $W(k)$ of \eqref{Bline2}.}
  \label{window}
\end{figure}

Before discussing constraints it is worthwhile pausing to consider exactly what 
quantities a particular observation might constrain. For example, FR observations
constrain the ``Rotation Measure'' which (in Minkowski space) is given by
\be
RM \equiv \frac{\Delta \phi}{\lambda^2} = \frac{e^3}{2\pi m_e^2} 
\int d{\bf l} \cdot \bfB \, n_e.
\label{RM}
\ee
where $\Delta\phi$ is the rotation angle of the linear polarization, $\lambda$ the
wavelength of the observed light, $n_e$ is the electron number density,
and the integration is along the line of sight. If the magnetic field is stochastic, observations
constrain the root-mean-squared value of $RM$, effectively constraining the ($n_e$
weighted) root-mean-squared line integral of the magnetic field. For illustrative purposes
we will consider $n_e$ to be uniform and constant and then we find,
\be
\la B_{\rm line}^2 \ra =\int_0^\infty dk \,  W(k) E_M(k)
\label{Bline1}
\ee
where $B_{\rm line}$ denotes the line integral of $\bfB$ along the line of sight, $\la\cdot \ra$ is 
an ensemble average, and $W(k)$ is a ``window function'', 
\be
W(k) = \frac{1}{2} \int_0^1 du\, \left ( 1 - u^2 \right ) {\rm sinc}^2 (\kappa u)
\label{Bline2}
\ee
where $\kappa \equiv kL/2$, $L$ is the distance to the source, and ${\rm sinc}(x)\equiv \sin(x)/x$. 
Fig.~\ref{window} shows a plot of $W(k)$. 
Therefore the constrained quantity is the convolution of the 
power spectrum and a window function that is specific to the observation. 
Since FR observations constrain an integral of $E_M(k)$, additional assumptions
are necessary to present a simple constraint on ``the magnetic field strength at
a certain length scale''. 

In cosmology the power spectrum might be expected
to be a power law that is either weighted on small length scales as in causal
scenarios, or weighted on large length scales as in inflationary scenarios.
For example, if $E_M(k) = E_{M*} (k/k_*)^n$ for $k \le k_*$ and $E_M(k)=0$
for $k > k_*$, then we can derive constraints on $E_{M*}$, $n$ and $k_*$. If
$n \le 0$, the integral in \eqref{Bline1} will be dominated by small $k$ and
can be done approximately to get the energy density: $\la B_{\rm line}^2 \ra \sim \rho_B$;
if $n$ is large, an approximate evaluation leads to a suppression of the integral by 
$k_* L$: $\la B_{\rm line}^2 \ra \sim \rho_B/(k_* L)$. This is to be expected since
the effect of small scale fields add up incoherently as in a random walk and the net 
FR is suppressed by a factor $1/\sqrt{N}$ where $N\sim k_*L$ is the number of steps in 
the random walk.

Alternately, we note that $W(k) \ge 0$ and $E_M(k) \ge 0$. Then a constraint such as
\be
\la B_{\rm line}^2 \ra = \int_0^\infty \frac{dk}{k} \,  W(k) k E_M(k) < B_*^2
\ee
means that the integral over every logarithmic interval also satisfies the constraint.
Then we can write the constraint as,
\be
B_\lambda^2 = 2k E_M(k) \lesssim \frac{2B_*^2}{W(k)}
\ee

Broadly speaking, current observations place upper bounds on the cosmological
magnetic field strength\footnote{The constraints are on the magnetic field in
CGS-Gaussian units. The conversion from Lorentz-Heaviside (LH) to CGS
magnetic field strength is $B^{\rm LH} = B^{\rm CGS}/\sqrt{4\pi}$.} 
$B_\lambda \lesssim 10^{-9}\,\rmG$ (up to ${\cal O}(1)$ factors)
for $\lambda \sim 100\,\Mpc - 1\,\Gpc$, while blazar cascade observations place 
lower bounds $B_\lambda \gtrsim 3\times 10^{-16}\,\rmG$ assuming
$\lambda \gtrsim 10\, \kpc$~\cite{Neronov73,Essey_2011,Finke_2015,
Biteau:2018tmv}\footnote{It should 
be noted that even though the existence of 
electromagnetic cascades is widely adopted, there is a possibility that plasma
instabilities may change the picture as we discuss in Sec.~\ref{blazars}.}.
We will refine this statement in Sec.~\ref{constraintplot}.

Here we will only discuss blazar observations in detail as these appear to have a lot 
of promise for detecting and measuring the magnetic field strength and also the
magnetic helicity. We will also briefly mention a recent promising proposal based on the 
effect of magnetic fields on cosmological recombination (see Sec.~\ref{Brecomb}).

\subsection{Electromagnetic cascades from blazars}
\label{blazars}

\begin{figure}
      \includegraphics[width=0.48\textwidth,angle=0]{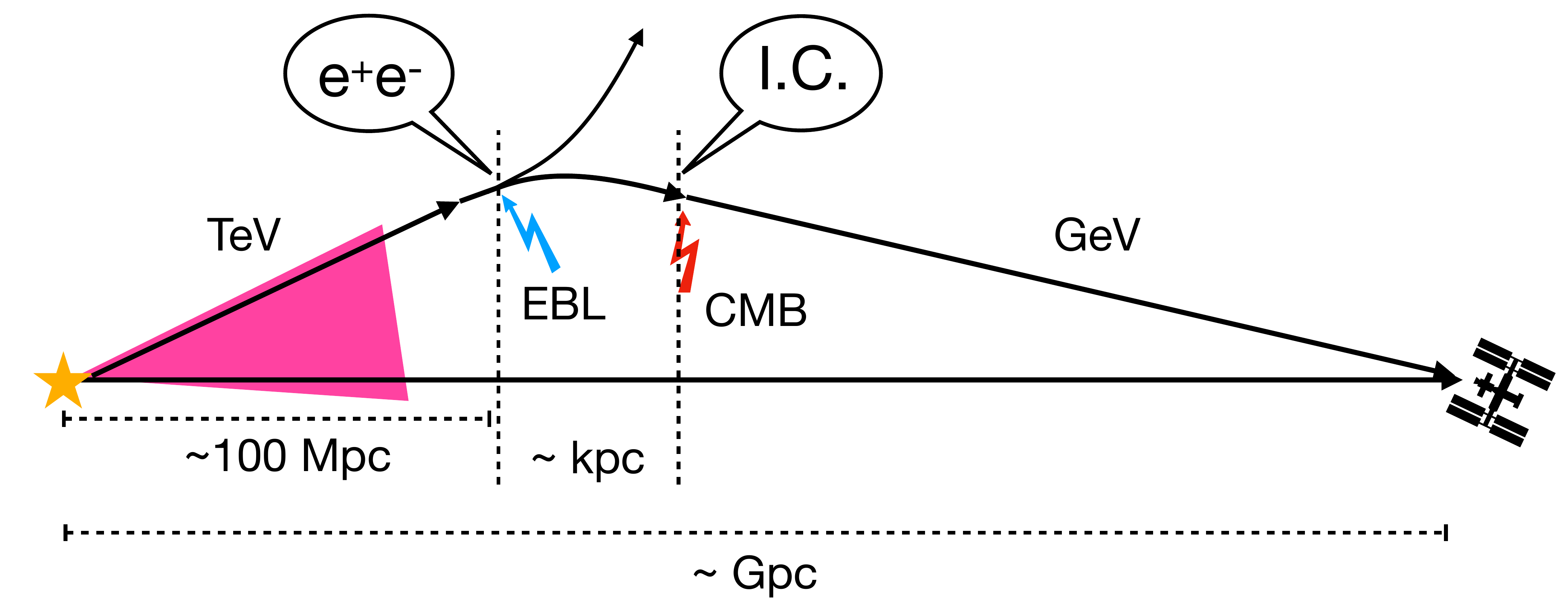}
  \caption{The three legs of the TeV photon's journey from the blazar to
  the observer in the presence of an inter-galactic magnetic field. The first
  TeV leg is terminated by pair production off an EBL photon, the second
  leg is in the form of lepton pairs and is terminated by an inverse
  Compton (I.C.) scattering event. The final GeV photon leg then propagates to
  the detector, perhaps onboard a satellite.
  Note that the second leg length scale of $\sim \kpc$ shown in the drawing
  is the mean free path of the electron and not the cooling distance which
  is $\sim 300\,\kpc$. In any case, the second leg of the journey is tiny
compared to the first leg $\sim 100~\Mpc$ and the third leg $\sim 1~\Gpc$. 
Only the second leg probes the inter-galactic magnetic field.}
  \label{blazarfig}
\end{figure}

The idea underlying the use of blazars to detect intergalactic magnetic fields 
is illustrated in Fig.~\ref{blazarfig}~\cite{Nikishov1961,PhysRevLett.16.252,d_Avezac_2007}.

Active galactic nuclei jets that are approximately pointed in our direction are 
called blazars. The jets have intrinsic opening angles $\sim 1^\circ$~\cite{Pushkarev_2017}
and they can emit very high energy gamma rays, including in the TeV energy range. There 
are 3 legs in the TeV photon's journey from the blazar to Earth as we now describe.

In the {\it first leg} of its journey, the TeV photon 
can encounter a photon of the ``extra-galactic background
light'' (EBL). 
The EBL is due to various sources of infrared photons in the universe, 
such as stars and active galactic nuclei (AGNs).
The EBL spectrum is not known with certainty but it
is modeled based on a variety of observations (see~\cite{Durrer:2013pga} for a summary). 
The EBL contains photons in the ultraviolet and optical, $0.1-10\,\eV$. This is important because
the TeV photon from the blazar can then scatter off an EBL photon and
the center of momentum energy will be above threshold to produce an
electron-positron (\eepair) pair. The distance that a TeV photon can 
travel before pair producing off the EBL is \cite{2009PhRvD..80l3012N,Durrer:2013pga},
\be
D (E)  \sim \frac{80~\Mpc}{(1+z_s)^2} \left ( \frac{10~\TeV}{E} \right ).
\ee
up to EBL model-dependent numerical factors; $z_s$ is the redshift of the
source.

The {\it second leg} of the journey is the propagation of the \eepair
that carry the original TeV energy. Kinematics tells us that
the angle the electron and positron make with the
forward direction is $\sim m_e/\TeV \sim 10^{-6}$ where $m_e = 0.5\,\MeV$ is
the mass of the electron.

Coming to the {\it third leg} of the journey, electrons and positrons
can only propagate a short distance before encountering
CMB photons -- the most abundant photons in the universe with energy
$\sim 10^{-4}\,\eV$ and number density
$n_{\rm CMB} \sim 500/\cm^3$. The mean free path is 
$l \sim 1/(n_{\rm CMB} \sigma_T) \sim \kpc$ where 
$\sigma_T = 6.6\times 10^{-25}\, \cm^2$ is the Thomson cross-section. 

Inverse-Compton scattering of the electron with a CMB 
photon~\cite{RevModPhys.42.237}, 
up-scatters the CMB photon to energy,
\be
E'_\gamma = \frac{4}{3} E_{\rm CMB} \frac{E_e^2}{m_e^2}.
\ee 
The resulting $\sim \GeV$ gamma ray has momentum within an angle
$m_e/E_e \sim 10^{-6}$
of the forward direction. 
In this process, 
the TeV electron loses a tiny fraction of its energy. Hence up-scattering
can continue to produce an ``electromagnetic cascade'' of 
GeV photons until the electron loses most of its energy, which  occurs
over a distance~\cite{2009PhRvD..80l3012N}
\be
D_{\rm IC} = \frac{3m_e^2}{4\sigma_T \rho_{\rm CMB} E_e} \approx
300 \left ( \frac{1\,\TeV}{E_e} \right ) \, \kpc
\label{DIC}
\ee
with $\rho_{\rm CMB} = 0.25 \, \eV /\cm^3$ being the energy density in the CMB.
Note that $300\, \kpc$ is much shorter than the $\sim 100\,\Mpc$ distance from the 
source as well as the $\sim \,\Gpc$ distance to the observer. It is only within this 
short distance that the propagation is via charged particles that are sensitive to the 
presence of a magnetic field. The resulting GeV gamma rays also have momenta within
an angle $\sim 10^{-6}$ with the forward direction.
Since the up-scattered CMB
photons now have GeV energies and are propagating very close to the forward
direction, they arrive to the observer as GeV gamma rays. Thus TeV
blazars should be observed to have GeV halos, also called ``pair halos'', and
the blazar spectrum should show an excess of GeV photons due to the cascade.

How do inter-galactic magnetic fields affect the electromagnetic cascade?
The trajectories of \eepair\, in the second leg of the journey get bent due to
the Lorentz force and, if the field is strong enough, the final GeV
photons are no longer directed towards the observer and the
GeV halo is not seen. Non-observation of 
the halo can lead to {\it lower} bounds on the strength of the inter-galactic
magnetic field (see Sec.~\ref{blazarspectra}); observations of a dispersed
halo can lead to measurements of the magnetic field strength 
(see Sec.~\ref{halodetection}).
An important caveat to this observational technique is that there be no 
other mechanism besides magnetic fields by which the
\eepair\, and the GeV photons can be dispersed (see Sec.~\ref{plasmainstability}).

To determine the spread of the halo due to a magnetic field we consider two
limiting cases. First, if the magnetic field is uniform on the scale of the Larmor radius,
$R_L = E_e/eB \sim 100\,\Mpc$ for $E_e\sim 1\,\TeV$ and $B \sim 10^{-17}\,\rmG$,
then the bending angle is $\sim D_{\rm IC}/R_L \sim 10^{-3} = 0.1^\circ$.
The other case, of more relevance to causal generation of magnetic fields, is if 
the magnetic field is isotropic on the scale of the Larmor radius, then the lepton 
trajectory is a random walk and the bending angle is \cite{2009PhRvD..80l3012N},
\ba
\delta &\approx& \frac{\sqrt{D_{\rm IC}\lambda_B}}{R_L} \nn \\
&\sim& 10^{-4} \left ( \frac{\lambda_B}{1\,\kpc} \right )^{1/2}
            \left ( \frac{B}{10^{-17}\,\rmG} \right ) \left ( \frac{1\, \TeV}{E_e} \right )
\ea
where $\lambda_B$ is the coherence scale of the magnetic field.
For $B \gtrsim 10^{-13}\, \rmG$ on kpc scales we find $\delta \sim 1$ and the small angle
approximation is not valid. We expect the halo to be very dispersed for such values of
the field strength.

 A novelty of using blazars to detect magnetic fields is that the technique
is immune to confounding effects at the blazar or in the Milky Way.
The \eepair\, only probe the magnetic field 
$\sim 100~\Mpc$ away from the blazar. Hence
this probe is insensitive to the processes occurring in the blazar. And
since the signal arrives at Earth in the form of GeV gamma rays, the
Milky Way magnetic field does not play any role. This is a tremendous
advantage of using blazar halos as an observational tool.

\subsubsection{Plasma instability?}
\label{plasmainstability}

The interpretation of missing blazar halos as being due to inter-galactic magnetic
fields has been debated in 
Refs.~\cite{Broderick_2012,2012ApJ...758..102S,2013ApJ...777...49S,2013ApJ...770...54M,Yan_2018,
pfrommer2013physics,Chang_2014,2014ApJ...783...96S,Chang_2016, Broderick_2016, 
Shalaby_2017,Tiede_2017,tiede2017constraints,Broderick_2018,Shalaby_2018, 
Batista_2019,Shalaby_2020}. The question is 
if the \eepair\, beam can excite a plasma instability and lose energy faster than the 
rate at which IC scattering cools the beam. If so, the energy in the \eepair\, beam
would go into heating the cosmological medium, not into GeV photons.

\begin{figure}
      \includegraphics[width=0.5\textwidth,angle=0]{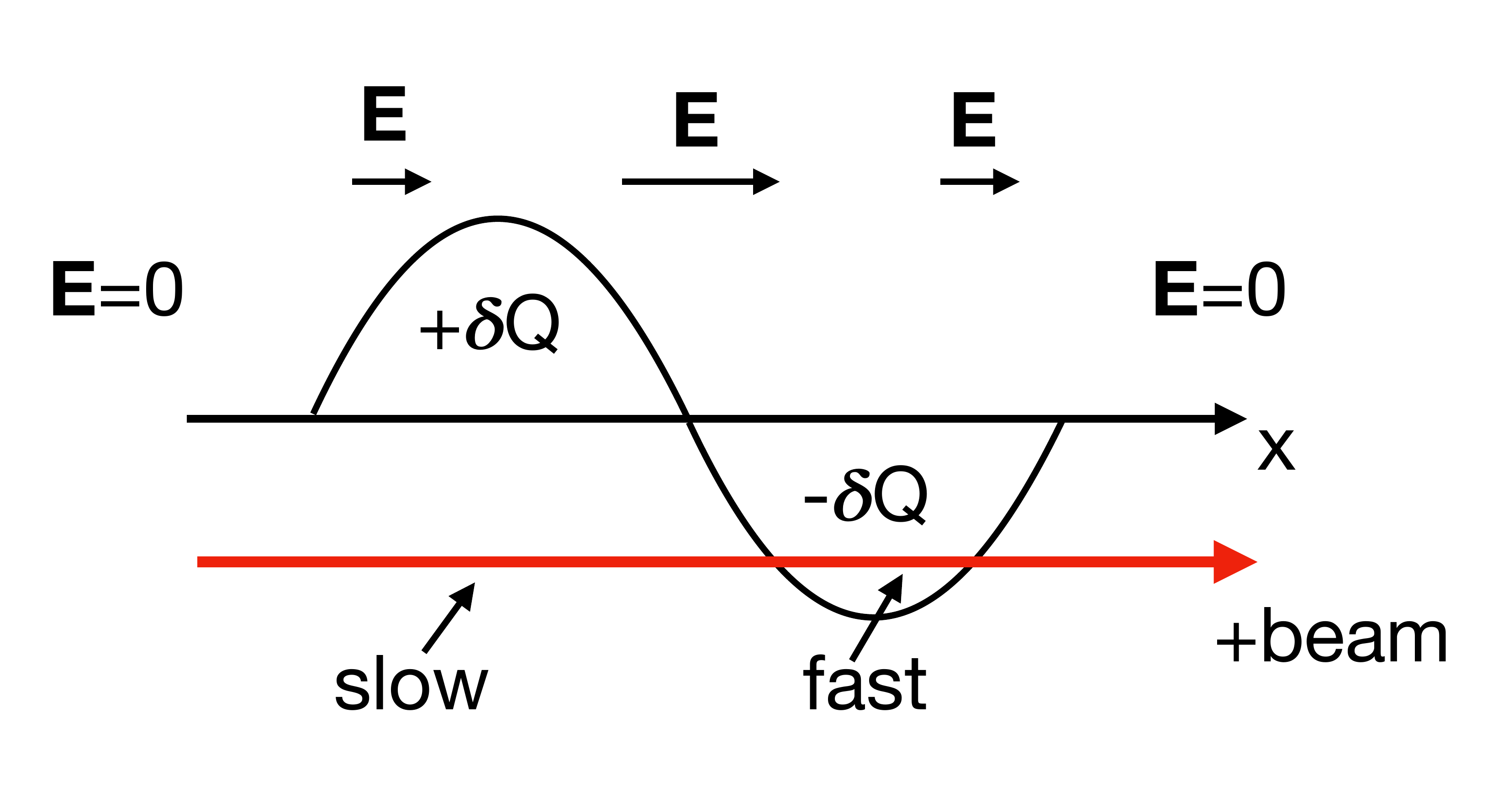}
  \caption{The drawing shows the basic physics underlying the plasma instability.
  A beam of charged particles (red arrow) is injected into a neutral plasma. The plasma may
  have perturbations in the charge density distribution so that there is an excess charge 
  $+\delta Q$ at one location and $-\delta Q$
  somewhere else, and these are taken to be a fixed background for illustrative
  purposes. (In general, charge perturbations would be expanded in modes.) The electric 
  field $\bfE$  due to this perturbation is depicted above and
  the key feature is that it points to the right everywhere (the $\pm \delta Q$ charge 
  distribution is like a charged capacitor) and the electric field is nearly zero away from 
  the perturbed region. Positively charged particles in the beam 
  enter from the left and are moving with their (``slow'') initial velocity as they enter the 
  $+\delta Q$ region where they get accelerated to the right by the electric field, and 
  then they move faster through the $-\delta Q$ region. Thus they spend more time
  in the $+\delta Q$ region than in the $-\delta Q$ region and effectively enhance 
  the perturbation. Similarly,
  negative charges in the beam spend more time in the $-\delta Q$ region and also
  enhance the perturbation. Other charged particles in the beam will effectively
  experience a larger charge perturbation and in this way the perturbation can grow. 
  This is the basic idea of the instability. In Ref.~\cite{Broderick:2011av} it is 
  argued that the instability grows fastest when the beam is obliquely incident on the 
  perturbation wavevector.}
  \label{instabilitycartoon}
\end{figure}

The basic physics of the main instability is illustrated in Fig.~\ref{instabilitycartoon}.
(Other instabilities, such as the Weibel instability, grow more slowly.)
The relevant comparison is the timescale for the growth of the instability versus the cooling 
time due to IC scattering. In Ref.~\cite{Broderick_2012}, the timescale of instability growth 
is estimated to be $10^3-10^5~{\rm yrs}$, whereas the IC
cooling distance estimated at $\sim 300\, \kpc$ in \eqref{DIC} equates to an IC cooling
time of $10^6\, {\rm yrs}$. Then the instability growth rate is faster and will be the main 
cause of the lepton beam dispersal. However, the matter is still under debate. Objections
have been raised based on simulation results, backreaction on the background, non-linear 
effects, and parameters such as the intrinsic spectrum of the blazar, density and temperature 
of the inter-galactic medium, and the luminosity of the beam (e.g.~\cite{Batista_2019}).

While the relevance of plasma instabilities is still under investigation, 
observations may be able to directly settle the issue -- if a dilute GeV halo is detected 
around TeV blazars then clearly plasma instabilities are not playing a role  since 
these dissipate the energy of the \eepair beam to heat and not to GeV photons
(see Sec.~\ref{halodetection} below).

\subsubsection{Evidence from blazar spectra}
\label{blazarspectra}

\begin{figure}
      \includegraphics[width=0.44\textwidth,angle=0]{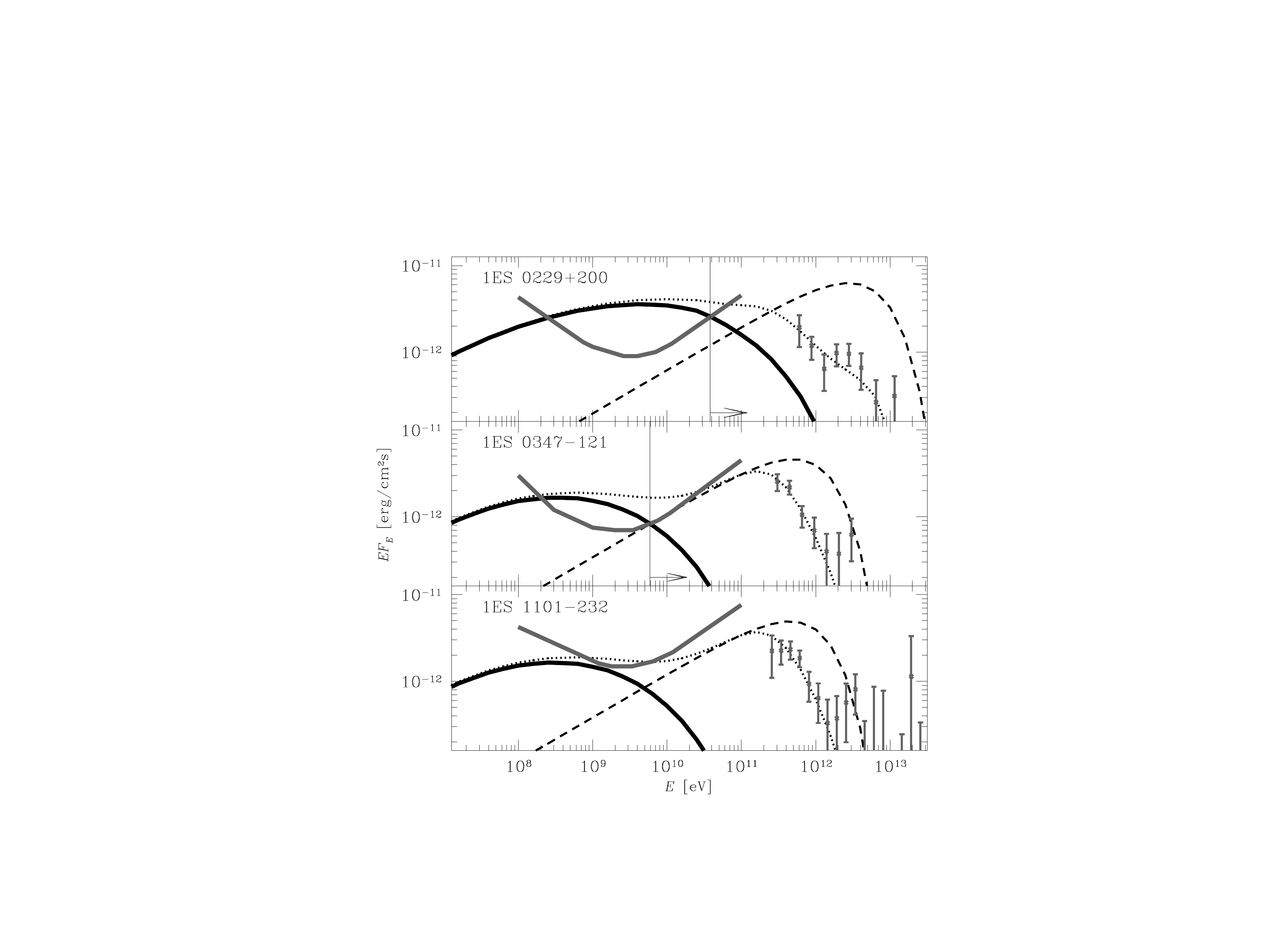}
  \caption{Data points from HESS observations for three blazars together with a 
  model of the EBL
  are used to deduce the primary source spectrum (dashed curves), the cascade
  contribution (solid curve) and the net processed  spectrum (dotted curves). 
  Constraints imposed by Fermi observations (grey concave solid curves) are 
  however 
  lower than the predicted spectrum at GeV energies. The deficit of GeV
  gamma rays provides lower bounds on the strength of the inter-galactic 
  magnetic field. [Plot from Ref.~\cite{Neronov73}.]}
  \label{neronovVovkPlot}
\end{figure}

An analysis of blazar spectra was performed in Ref.~\cite{Neronov73} and
obtained a lower bound $\sim 10^{-16}~\rmG$ on the inter-galactic
field assuming homogeneity on 1~Mpc scales. Since then a number of other
groups have performed similar analyses, though the lower bounds have varied
from $\sim 10^{-19}~\rmG$ to $\sim 10^{-17}~\rmG$ depending on the details
of the analysis~\cite{Neronov73,Essey_2011,Finke_2015,Biteau:2018tmv}.
The analysis of \cite{Arlen:2012iy} concludes that the data is consistent with 
zero magnetic field but may have adopted an unrealistic model for the intrinsic
spectra of blazars and the EBL~\cite{Durrer:2003ja}.
Fig.~\ref{neronovVovkPlot} shows the 
main plot from Ref.~\cite{Neronov73} for the spectra from three blazars and the 
analysis. 

The absence of GeV halos can be explained by magnetic fields that are
stronger than $10^{-16}\,\rmG$ on $\sim\Mpc$ scales as in~\cite{Neronov73}
but other combinations of strength and coherence scale are also possible.
For example, magnetic fields weaker than $10^{-16}\,\rmG$ on $\sim\Mpc$ scales
but stronger on smaller scales, \iteg~$10^{-14}\,\rmG$ on $\sim\kpc$ scales,
can equally well cause sufficient deflection of the lepton pairs and dilute
the GeV halos. In other words, the lower bound is on an
appropriately weighted integral of the power spectrum $E_M(k)$ and
the quoted bound of ``$10^{-16}\,\rmG$ on $\sim\Mpc$ scales'' should
not be taken too literally.

The important point of this analysis is that there exists a {\it lower} bound on the 
magnetic field strength, with the most recent analysis by the Fermi
collaboration~\cite{Biteau:2018tmv},
even if the numerical value of the lower bound is model-dependent. 
The only known alternative explanation for the spectral signature is the plasma instability 
discussed in Sec.~\ref{plasmainstability}.
Yet it may be possible to distinguish between the two scenarios by direct
observations of the halo as we now discuss.

\subsubsection{Detection of the halo}
\label{halodetection}

\begin{figure}
      \includegraphics[width=0.5\textwidth,angle=0]{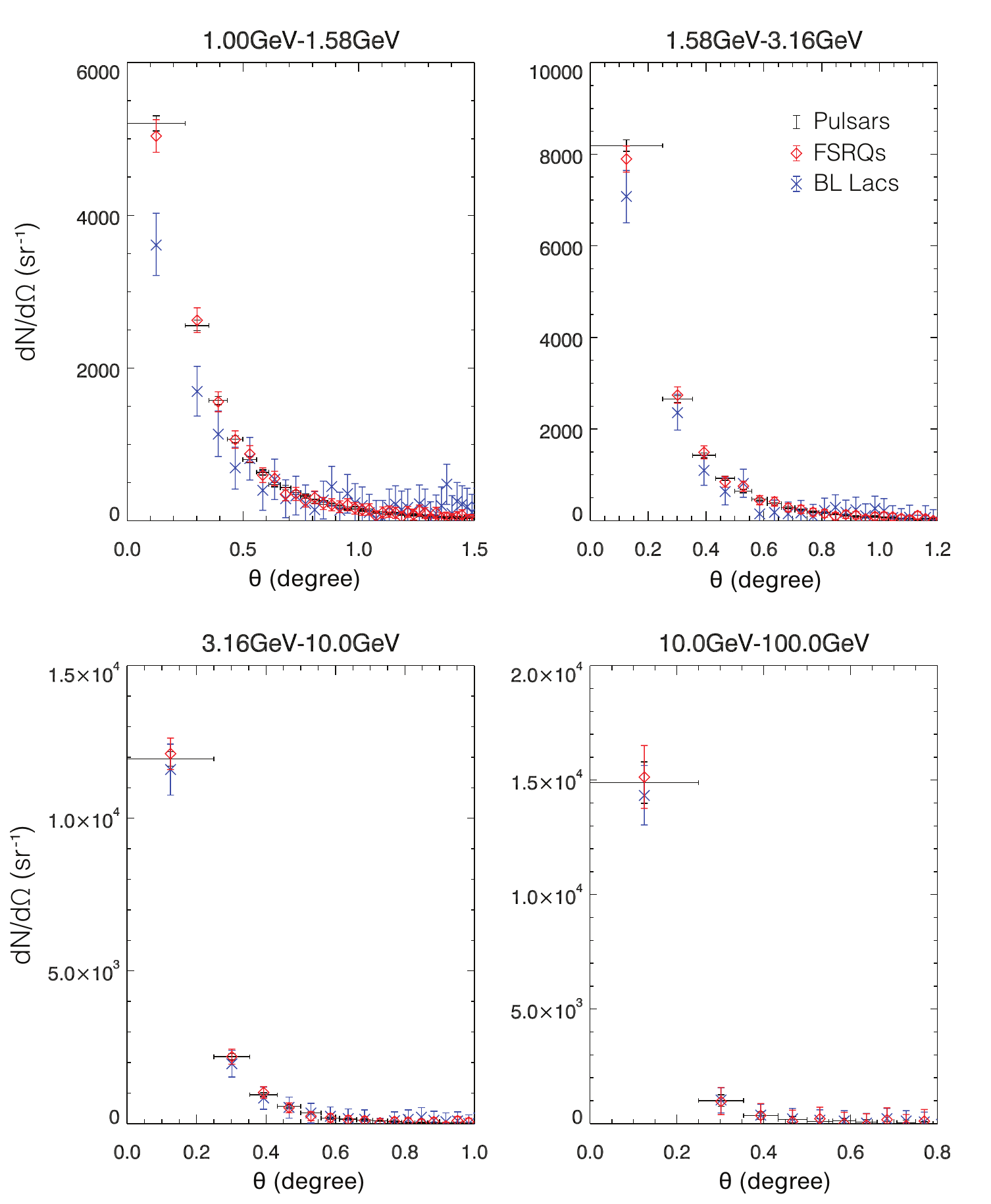}
  \caption{Data points show the count of GeV photons in various energy bins versus 
  angular distance from 24 stacked sources for 3 different classes of sources:
  Pulsars (black bar), FSRQs (red bar with circle), and the TeV-luminous BL Lacs
  (blue bar with cross). The pulsar and FSRQ counts 
  agree with each other but there is an excess count from the BL Lacs at large angular
  distances with $\sim 3.5\sigma$ significance in the lowest energy bin (top left plot)
  as would be expected if there was a pair halo around BL Lacs.
  [Plot from Ref.~\cite{Chen:2014rsa}.]}
  \label{bllacsvsfsrqs}
\end{figure}

The pair halo around a single blazar may be too diffuse to be seen with current
observations. The strategy employed in 
Refs.~\cite{2010ApJ...722L..39A,Chen:2014rsa} is to stack images of 
several different blazars to get a larger photon count and then look for an excess 
of pair halo GeV photons. 
To decide if there is an excess one compares
the stacked results for TeV sources (BL Lacs) with an identical stacked 
analysis for similar sources but for which pair halos are not to be expected
(Flat Spectrum Radio Quasars or FSRQs). 
In Fig.~\ref{bllacsvsfsrqs} we show the results from Ref.~\cite{Chen:2014rsa}.

The claimed detection of pair halos is not universally accepted.
The earliest claimed evidence for pair halos in a stacked analysis~\cite{2010ApJ...722L..39A} 
was argued to be due to an instrumental effect~\cite{neronovetal2011}.
Other analyses by the Fermi collaboration~\cite{Biteau:2018tmv}
and the VERITAS collaboration~\cite{Archambault_2017} did not corroborate 
evidence for pair halos. Further data will help 
to clarify the situation; improved analysis techniques, such as the idea to use radio 
observations to align blazar jet directions prior to stacking 
can enhance the sensitivity of the stacking method~\cite{Tiede:2017xng,Chen:2018mjd}.

\subsubsection{Halo morphology}
\label{halomorphology}

Several groups have studied the shape and structure of pair 
halos~\cite{Elyiv:2009bx,AlvesBatista:2016urk,AlvesBatista:2017jga,
Long:2015bda,Duplessis:2017rde,Broderick:2016akd, Fitoussi:2017ble,kachelriess2020searching}.
A full simulation is complicated, especially in a stochastic
magnetic field, but some features of the effect of a magnetic field on the
halo are simple to understand based on the structure of the
surface where pair production can occur~\cite{Duplessis:2017rde}.

\begin{figure}
      \includegraphics[width=0.4\textwidth,angle=0]{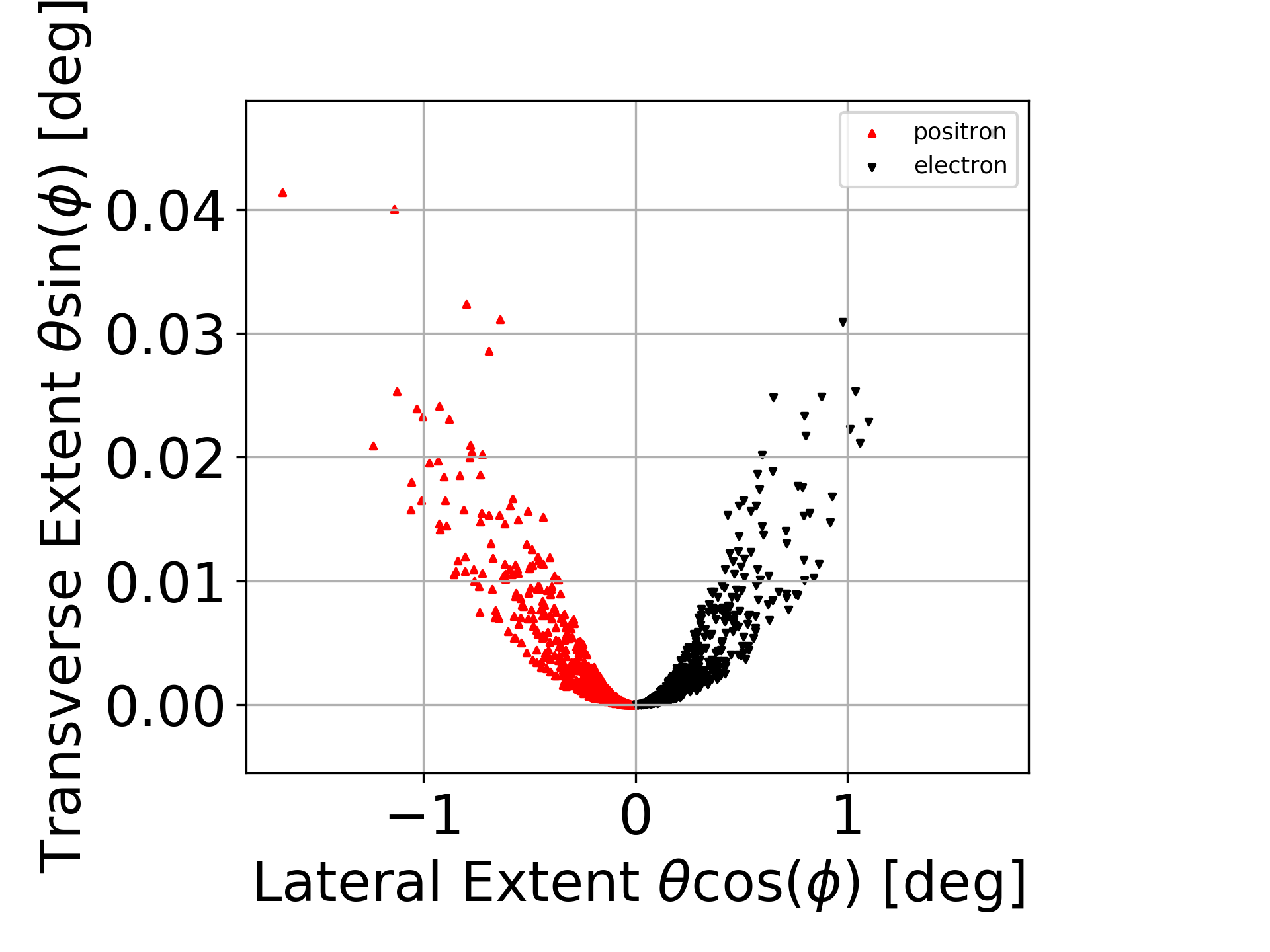}
      \caption{The halo for a magnetic field that is at a $45^\circ$ angle to the line
      of sight to a blazar whose jet is pointed to the observer. (Photon energy information 
      is not shown.) If the magnetic
      field was chosen to be orthogonal to the line of sight, the two branches would
      have been straight, giving a bow-tie shape. The red dots correspond to photons
      that have been up-scattered by positrons and the black dots due to 
      up-scattering by electrons. If the blazar jet axis is misaligned with the line of
      sight, the bow-tie will not be symmetrical.
      }
  \label{francis1}
\end{figure}

\begin{figure}
      \includegraphics[width=0.4\textwidth,angle=0]{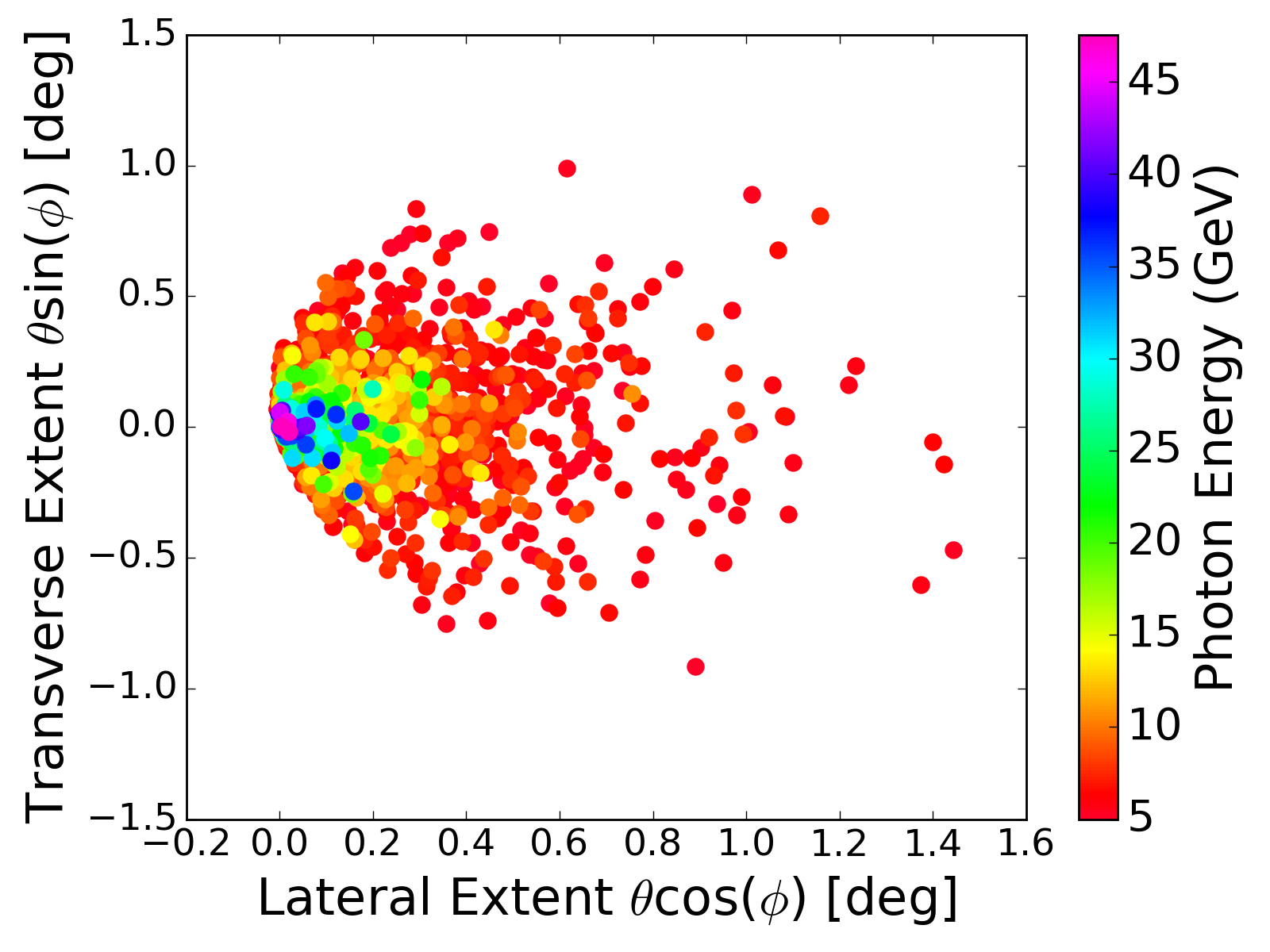}
      \caption{The halo for a helical field and a blazar jet that is misaligned with the
      line of sight. (Details may be found in Ref.~\cite{Duplessis:2017rde}.) The
      misalignment in this case eliminates one of the lepton branches of the halo.
      The different color dots represent photons of various energies as shown on the
      bar on the right and the energetic photons lie closer to the line of sight. 
      The helicity of the magnetic field provides a twist to the 
      distribution of photons of different energies in the halo.}
  \label{francis2}
\end{figure}

\begin{figure}
      \includegraphics[width=0.4\textwidth,angle=0]{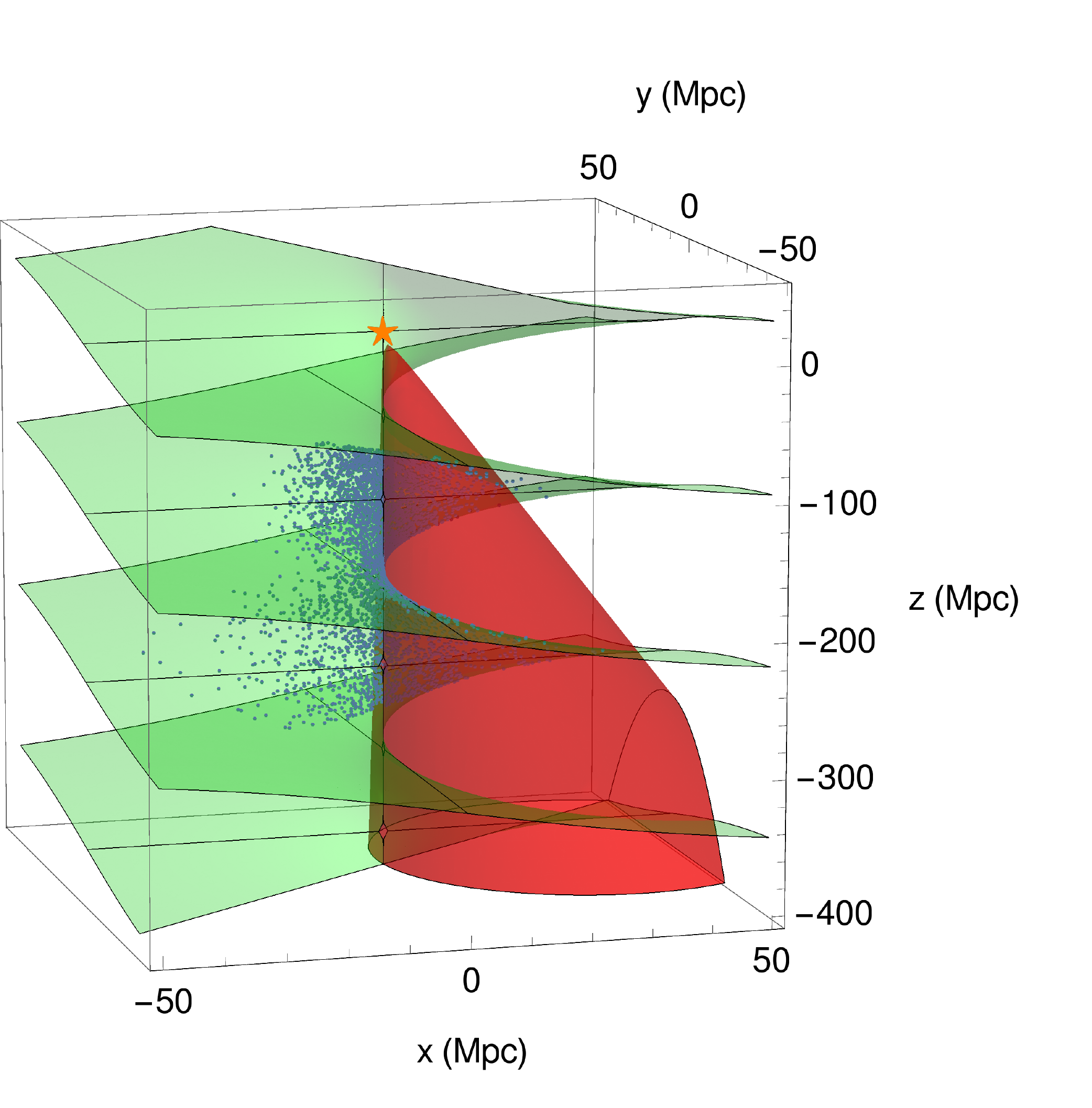}
      \caption{The TeV photons from the blazar (yellow star) travel a certain distance, 
      pair produce,
      up-scatter CMB photons that reach the observer. This process maps out a 
      spatial surface called the ``particle production surface'' (shown in green) that 
      depends on the inter-galactic magnetic field, such that 
      only pair creation events on the surface can potentially be
      observed. For helical magnetic fields, the surface has a spiral structure.
      The blue dots represent potential pair creation events that can be observed though
      an actual jet (shown as the red cone) may not activate all the possible observable
      pair creation events~\cite{Duplessis:2017rde}.}
  \label{francis3}
\end{figure}

Consider the simple setup where the blazar jet points directly at
the observer and there is no magnetic field ($B=0$). Then, assuming
an axially symmetric jet, the halo will also be axially symmetric and will
appear as a disk to the observer. Next let us introduce a uniform magnetic 
field that is orthogonal to the line of sight. This breaks the axial symmetry
because charges bend by different amounts depending on their direction
of propagation and the sign of the charge~\cite{Long:2015bda}. Then 
the halo stretches out in the two directions orthogonal to 
the line of sight and to the magnetic field direction. The higher
energy charged particles are bent less by the magnetic fields and the
gamma ray cascade they produce tend to cluster close to the
line of sight, while the lower energy gamma rays lie further away
from the line of sight. This gives the halo a ``bow-tie'' 
structure~\cite{Broderick:2016akd}, at least for magnetic 
fields that are highly coherent, as illustrated in Fig.~\ref{francis1}. The two sides of 
the bow tie also have
an interesting origin. When the TeV gamma rays pair produce, the
electrons will bend one way due to the Lorentz force exerted by the magnetic field 
while the positrons
will bend in the opposite way. One side of the bow-tie is due to photons that
have been up-scattered by electrons, while the other side of the bow-tie
is due to up-scattering by positrons.
If the blazar
jet is not directly pointed at the observer, one side of the bow-tie will be
less prominent than the other as in Fig.~\ref{francis2}.
If the magnetic field is helical, the bow-tie shape also gets twisted,
either in a clockwise or in a counter-clockwise direction depending
on the sign of the helicity, also seen in Fig.~\ref{francis2}). Features
of the halo may be understood in terms of a ``particle production surface''
as in Fig.~\ref{francis3}. 


\subsubsection{Search for magnetic helicity}
\label{searchforhelicity}

The twisting of the halo provides a handle for the detection of inter-galactic 
magnetic fields and their helicity. This might seem like a ``second order effect'' 
-- we first need to detect magnetic fields which is hard and then its helicity which
seems harder. However, since helicity is a parity odd feature, its signatures are
protected from confusion with foregrounds  and most other sources of noise as
those are generally parity even. 
There is an additional high-stakes reason for seeking magnetic helicity as
it would be direct evidence for the violation of fundamental symmetries (P, CP)
in particle physics and cosmology.
On top of this, a detection of magnetic helicity will also resolve the ambiguity 
between the magnetic field interpretation of missing pair halos and the interpretation 
in terms of plasma instabilities. Another motivation to search for magnetic helicity
is that MHD evolution, to be discussed later, shows that helicity is an essential feature 
if causally generated magnetic fields are to survive on large length scales. If magnetic
fields are discovered but they are not helical, it would indicate an acausal
generation mechanism or an astrophysical mechanism in the
very recent universe.

The detection of magnetic helicity using parity odd correlators of the 
CMB was discussed in Ref.~\cite{Caprini:2003vc}.
Techniques for detecting magnetic helicity using blazar data were first discussed 
in~\cite{Tashiro:2013bxa,Tashiro:2014gfa} following ideas in~\cite{Kahniashvili:2005yp}.
If ${\hat l}$ denotes the line of sight to a blazar, the twisting of a particular halo can be 
measured by $ {\hat n}_1 \times {\hat n}_2 \cdot {\hat l}$ 
where ${\hat n}_a$ refers to the arrival directions of gamma rays of energy $E_a$,
and we order the energies so that $E_1 < E_2$.
Hence the twisting of a single blazar halo can be measured by computing
\be
Q(R) = \la {\hat n}_1 \times {\hat n}_2 \cdot {\hat l} \ra_R
\ee
where the angular brackets refer to an average over all photons around the 
blazar out to an angular distance of $R$.

In fact, the line of sight direction nearly coincides with the direction of propagation
of the highest energy gamma rays. So the line of sight direction can be approximated
by ${\hat n}_3$ for a photon with very high energy $E_3$ and the twisting can be measured
by calculating ${\hat n}_1 \times {\hat n}_2 \cdot {\hat n}_3$ where 
the energies are ordered: $E_1 < E_2 < E_3$. Since now there is
no reference to an observed blazar, this quantity can be calculated
over the entire sky with sums over all photons with energies $E_1$,
$E_2$ and $E_3$. This will tell us if there is an overall handedness in
the gamma ray sky. 
This leads to the statistic $Q$ for measuring magnetic helicity over the
entire sky whether or not blazars are identified,
\be
Q({\hat n}_1,{\hat n}_2, {\hat n}_3; R) =
\frac{1}{N} \sum_{ \{ n_a \} }   
{\hat n}_1 \times {\hat n}_2 \cdot {\hat n}_3
\label{Qdefn}
\ee
where $N$ is the total number of terms in the sum.
Since $Q$ is parity odd, it is no surprise that it is proportional to the helical
part of the magnetic field correlator, $M_H$, in \eqref{BBcorr}. In the 
idealized case of no backgrounds, small deflection angles, and large
angular regions (to capture all GeV photons)~\cite{Tashiro:2014gfa},
\be
Q(E_1,E_2,E_3;R\to \pi/2) \propto r \, M_H(r)
\ee
where the distance $r$ is related to the energies $E_1$ and $E_2$. 
By considering different photon energies $E_1$ and $E_2$, the statistics
$Q$ can probe the helical spectrum at different spatial separations and, in 
principle, we can recover the entire helical power spectrum. As we shall see in
Sec.~\ref{evolution}, causally generated magnetic fields are expected to be
maximally helical, in which case the helical power spectrum is related
to the power spectrum, $E_M(k)$. Thus we can recover the entire correlation
function for the magnetic field from the $Q$ statistics.

\begin{figure}
      \includegraphics[width=0.45\textwidth,angle=0]{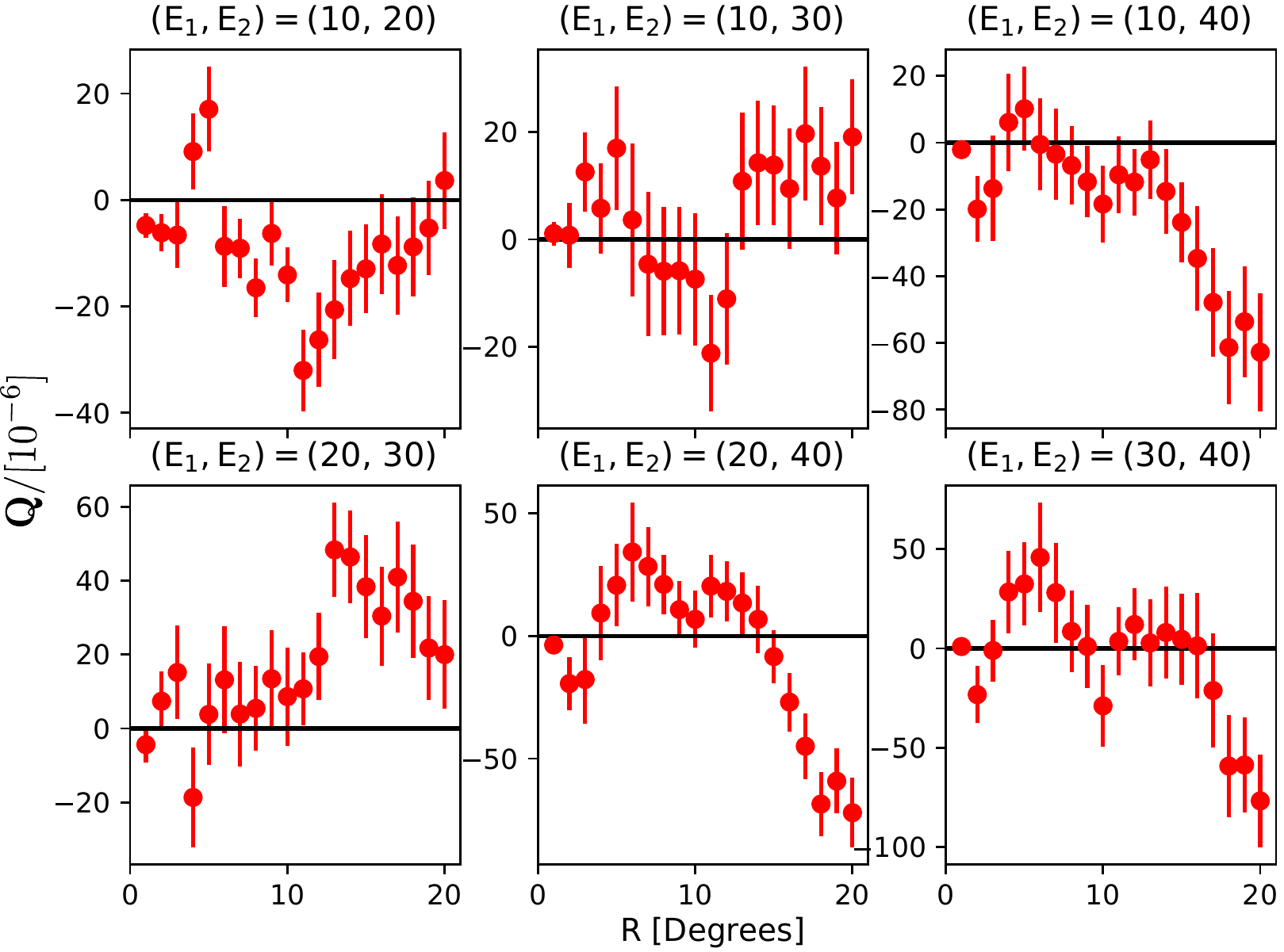}
  \caption{The $Q/10^{-6}$ statistic plotted vs. the angular size $R$ for various values of
  $(E_1,E_2)$ and for $E_3 > 50\,\GeV$. These plots are for 11 years of Fermi data,
  following the scheme in~\cite{Tashiro_2014,Chen_2015}. [Plot by Teerthal Patel, unpublished.]}
  \label{QvsR}
\end{figure}

To actually compute $Q$ from data requires some more refinements as there 
are confounding gamma rays that originate in the Milky Way and in other known
sources. Using 11 years of Fermi Observatory data and only including the highest 
energy gamma rays ($E_3 =50-60\,\GeV$) at galactic latitudes above $80^\circ$,
and for various combinations of $E_1$ and $E_2$ leads to
the $Q(R)$ plots shown in Fig.~\ref{QvsR}. The error 
bars in the plot are statistical. The question is if the non-zero values of $Q$ are
significant. This requires a careful analysis such as in 
Refs.~\cite{asplund2020measurement,kachelriess2020searching} with the conclusion 
that the plots of $Q$ are consistent with vanishing helicity.

At this point in time the detection of magnetic helicity via the twisting of halos is still
not conclusive but, given the compelling motivations for pursuing helicity,
it is important to develop new strategies. Already a refinement of $Q$ has
been proposed in~\cite{Duplessis:2017rde} that limits the sum in \eqref{Qdefn}
to just one branch of the bow-tie (to avoid terms that involve cross products of
GeV photon direction vectors from electron up-scatterings and those from positron 
up-scatterings). 
In relatively simple simulations the refined $Q$
is seen to have greater sensitivity to magnetic helicity~\cite{Duplessis:2017rde} . 
On the observational
front, in future we expect more gamma ray data will be available around individual blazars 
on which to apply the $Q$ statistic. Or perhaps the evaluation of $Q$ on stacked 
data will be more conclusive. Here we would need to ensure that the halo bow
ties are aligned prior to stacking as mentioned in Sec.~\ref{halodetection}.

\subsubsection{Pair echoes}
\label{pairechoes}

If a transient source such as a GRB emits TeV photons,
these photons too will undergo a cascade and produce GeV photons as
described above. In the presence
of inter-galactic magnetic fields, the GeV photons will be delayed because 
the path length after bending is longer than the line of sight path length
(see Fig.~\ref{blazarfig}) and lead to a ``pair echo''.
Stronger magnetic fields will cause greater time delays in the arrival of
halo photons of a given energy. Therefore a measured time delay can
lead to a measurement of the magnetic field strength, while non-observation
of an echo can lead to a lower bound~\cite{Plaga:1995ins,Taylor_2011,Veres_2017}.
Ref.~\cite{Taylor_2011} finds a bound $\gtrsim 10^{-17}\,\rmG$, while
\cite{Veres_2017,Wang:2020vyu} obtain $\gtrsim 10^{-19}\,\rmG$.
The analysis of~\cite{Wang:2020vyu} has been criticized 
in~\cite{Dzhatdoev:2020yvn}.

\subsection{Observations at high redshift}

\subsubsection{Magnetic fields in high redshift galaxies}

The Milky Way has magnetic field strength $\sim 10^{-6}\,\rmG$, 
rotational velocity $\sim 200 \,{\rm km/s}$, and disk radius 
$\sim 30\, \kpc$ and baryon density $\rho_{\rm gal} \approx 10^{-24}\, {\rm gm/cm^3}$.
If we assume flux freezing during gravitational collapse of a proto-galaxy, the 
magnetic field strength increases in this process by a factor
$(\rho_{\rm gal}/\rho_b )^{2/3} \sim 10^5$ where $\rho_b \sim 10^{-31}\, {\rm gm/cm^3}$ 
is the average cosmological baryon density. Therefore $\sim 10^{-11}\, \rmG$
cosmological magnetic fields can give rise to the galactic field strength $\sim \mu\rmG$
simply by gravitational compression. 

Assuming that the observed galactic field of $\sim 1\,\muG$  is due to dynamo amplification 
of a seed field in addition to gravitational compression, and each full rotation of the 
galaxy amplifies the field strength by an e-fold (``a maximally efficient dynamo''),
the seed field required to explain the Milky Way magnetic field is
$\sim e^{-40} 10^{-5}10^{-6}~\rmG$ or about $10^{-29}~\rmG$, since the Milky
Way is estimated to have made $~40$ complete revolutions. Such a tiny
seed field may possibly have arisen from a Biermann battery
mechanism assuming certain dynamics during galaxy 
formation~\cite{1994MNRAS.271L..15S,Kulsrud:1996km}.

A difficulty with the large-scale dynamo is that micro Gauss fields have also been 
observed in galaxies at a redshift $\sim 2$~\cite{Bernet:2008qp,Kronberg:2007dy}
when the universe was only $\sim 1/5$ of its present age.
Such galaxies would at most have gone through
$\sim 40/5=8$ revolutions and the maximum dynamo amplification would only be by
a factor $\sim 10^3$. Then the seed field would need to be around a $10^{-14}\,\rmG$
which is much harder to arrange by astrophysics alone. An alternate scenario based
on a turbulent dynamo is discussed in \cite{Kulsrud:1996km}.

\subsubsection{Magnetic fields and cosmological recombination}
\label{Brecomb}

CMB anisotropies generally impose constraints on cosmological magnetic
fields at the nano Gauss level. To see that this is reasonable, consider that
CMB observables mostly depend on adiabatic density fluctuations in the primordial 
plasma with $\delta\rho/\rho \sim 10^{-5}$.
As estimated in Sec.~\ref{somenumbers}, the energy density in micro Gauss magnetic 
fields is of the same order as that in CMB photons today and so nano Gauss 
magnetic fields can cause density fluctuations on the order of $10^{-6}$. Stronger
magnetic fields can start interfering with the successful predictions of adiabatic density
fluctuations. This shows very roughly that CMB observations can be expected to 
lead to constraints on magnetic fields at the nano Gauss level.

As discussed above, most cosmological observations are sensitive to large-scale 
magnetic fields
but small-scale, say kpc, magnetic fields remain weakly constrained. Since causally
generated magnetic fields are peaked on small length scales, it is important to find
ways to observe and constrain them. The CMB may
be used to constrain small-scale fields, not by the large scale anisotropies, instead by 
using spectral observations (departures from the blackbody spectrum).
The effect of small-scale magnetic fields on spectral distortions of the CMB have 
been considered in Refs.~\cite{Jedamzik:1999bm,Kunze:2014eka,Wagstaff:2015jaa}
and generally also lead to nano Gauss constraints on kpc-Mpc scales.

Recently another effect of magnetic fields on recombination and the CMB has been 
discussed by Jedamzik and 
collaborators~\cite{Jedamzik:2013gua,Jedamzik:2018itu,Jedamzik:2020krr}. 
The basic idea is that the MHD equation for the plasma flow contains the Lorentz 
force term $\propto \bfB\times ( \nabla \times \bfB )$ (see \eqref{navstokes}). The 
stochastic magnetic field will therefore create stochastic density fluctuations in the 
baryon fluid density~\cite{Jedamzik:2013gua,Jedamzik:2018itu} that
depend mainly on the magnetic field power spectrum on the smallest length scales. 
The Hydrogen recombination rate
is proportional to the square of the baryon density and since the average of the
square, $\la n_b^2 \ra$, is always larger than the square of the average,
$\la n_b \ra^2$, baryonic inhomogeneities lead to faster recombination overall
and the sound horizon at recombination, $r_s$, is smaller due to the magnetic field.
Temperature fluctuation correlations of the CMB 
are measured as a function of the angular separation 
of points on the sky and the dominant peak in the spectrum is at an angle,
$\theta = r_s/l \approx 1^\circ$ where $l$ is the distance to the surface of recombination.
Hence a smaller sound horizon implies that $l$ is also smaller. 
The Hubble constant $H_0$ enters the expressions for both $r_s$ and $l$ but the
dependence of $l$ on $H_0$ is stronger~\cite{pogosian2020recombinationindependent}.
Since $l$ decreases with increasing $H_0$ while $r_s$ stays approximately constant, we
need a larger $H_0$ to keep $\theta \approx 1^\circ$ in the presence of magnetic fields.
If we do not take the magnetic field into account, CMB measurements
will give an anomalously low value  of the Hubble constant that will differ from $H_0$ 
measurements made at lower redshifts. In this way, primordial magnetic fields 
of strength $\sim 0.1\,{\rm nG}$ on kpc scales~\cite{Jedamzik:2020krr} can alleviate
some of the current tension in the supernovae (nearby) measurements of 
the Hubble constant, $H_0= 73.5\pm 1.4~{\rm km/s/Mpc}$~\cite{Reid:2019tiq}, 
and the (distant) measurements using the CMB
that so far have been giving $H_0 = 67.37\pm 0.54~{\rm km/s/Mpc}$~\cite{Aghanim:2018eyx}.
Such magnetic fields would comfortably satisfy other existing constraints especially if the 
spectrum is peaked on short scales like that of causally generated fields. Furthermore, 
the field strength is in the range where it can directly source observed galactic and cluster 
magnetic fields with minimal dynamo processing and would explain the observation
of magnetic fields in high redshift galaxies.

Although still young, this idea is exciting not only because it potentially resolves 
a crisis in cosmological observations (the ``Hubble tension'') but also because it 
would be the earliest observational signature of cosmological magnetic fields. 
It would, by itself, point to an early universe generation mechanism of magnetic 
fields and indicate new particle physics and cosmology as we discuss in Sec.~\ref{production}.

\section{Production of magnetic fields in the early universe}
\label{production}

Several ideas for the generation of magnetic fields in the early universe have
been proposed based on particular epochs in cosmology, such as inflation, the 
electroweak phase transition, the QCD phase transition, and recombination. 
Here I will focus on magnetic fields generated at the electroweak phase
transition and will only make brief remarks on the other possibilities in
Sec.~\ref{otherideas}.

\subsection{Magnetic fields from the electroweak phase transition}
\label{ewptB}

\subsubsection{Defining the electromagnetic field strength}
\label{emdefinition}

The electroweak model contains four gauge fields corresponding to the
three generators of the ``weak'' $SU(2)$ and one generator of hypercharge 
$U(1)_Y$. The $SU(2)$ gauge fields are usually denoted by $W_\mu^a$ ($a=1,2,3$ 
and $\mu$ is the Lorentz index) and the hypercharge gauge field by $Y_\mu$.
The vacuum expectation value (VEV) of the $SU(2)$ doublet Higgs field
$\Phi$, breaks 
the electroweak $SU(2)\times U(1)_Y$ symmetry to the electromagnetic $U(1)_Q$.
The electromagnetic gauge field, $A_\mu$, is a linear combination of the weak
and hypercharge gauge fields,
\be
A_\mu = \sin\theta_w {\hat n}^a W^a_\mu + \cos\theta_w Y_\mu ,
\label{Adefn}
\ee
where $\theta_w$ is the weak mixing angle ($\sin^2\theta_w \approx 0.23$),
\be
{\hat n}^a \equiv - \frac{\Phi^\dag \sigma^a \Phi}{\Phi^\dag \Phi}
\ee
and $\sigma^a$ are the Pauli spin matrices 
\be
\sigma^1=\begin{pmatrix} 0 & 1 \\ 1 & 0 \end{pmatrix}, \ \ 
\sigma^2=\begin{pmatrix} 0 & -i \\ i & 0 \end{pmatrix}, \ \ 
\sigma^3=\begin{pmatrix} 1 & 0 \\ 0 & -1 \end{pmatrix}
\ee
The vector ${\hat n}^a$ is ill-defined at locations where $\Phi=0$. These points may
correspond to locations of magnetic monopoles and electroweak strings. However, at 
late times $\Phi$ relaxes to the true vacuum with $|\Phi | = \eta \approx 174\, \GeV$
everywhere and then $n^a$ is globally defined.

One might think that the electromagnetic field strength, $A_{\mu\nu}$,
equivalently the electric and magnetic fields, should be defined in the
usual way
\be
A_{\mu\nu} \stackrel{?}{=} \sin\theta_w {\hat n}^a W^a_{\mu\nu} + \cos\theta_w Y_{\mu\nu}
\label{Amunuguess}
\ee
but this definition has the difficulty that the right-hand side does not
correspond to the Maxellian $\partial_\mu A_\nu - \partial_\nu A_\mu$ for two 
reasons: first, the derivatives $\partial_\mu A_\nu$ will also in general act on ${\hat n}^a$;
second, $W^a_{\mu\nu}$ has terms that are quadratic in the gauge fields
$W^a_\mu$  and these are absent in $\partial_\mu A_\nu -\partial_\nu A_\mu$
as calculated from the definition in \eqref{Adefn}. The former difficulty can be avoided in 
the broken phase where $\Phi$ can be gauge transformed to a constant (the ``unitary gauge'') but
we are still stuck with the second difficulty. We would like to eliminate
the quadratic term in $W^a_\mu$ in \eqref{Amunuguess} while obtaining
the Maxwellian field strength in the unitary gauge (in which $\Phi$ is uniform). 
Following 't~Hooft's definition for an SO(3) model~\cite{tHooft:1974kcl},
a resolution is to define~\cite{Vachaspati:1991nm}
\ba
A_{\mu\nu} &=&  \sin\theta_w {\hat n}^a W^a_{\mu\nu} + \cos\theta_w Y_{\mu\nu} \nn \\
&&
 - i \frac{2\sin\theta_w}{g\eta^2} ( D_\mu\Phi^\dag D_\nu\Phi - D_\nu\Phi^\dag D_\mu\Phi )
\label{Amunu1}
\ea
in the true vacuum where $\Phi^\dag \Phi = \eta^2$.
A little algebra then shows that
\ba
A_{\mu\nu} &=& \partial_\mu A_\nu - \partial_\nu A_\mu \nn \\
&&
 - i \frac{2\sin\theta_w}{g\eta^2} 
 ( \partial_\mu\Phi^\dag \partial_\nu\Phi - \partial_\nu\Phi^\dag \partial_\mu\Phi )
\label{Amunu2}
\ea
Then there can be a non-zero electromagnetic field even if $A_\mu=0$ due to 
gradients of the Higgs field\footnote{It may be remarked that this is exactly the
situation for the 't~Hooft-Polyakov magnetic monopole in the hedgehog 
gauge where the magnetic field is precisely due to the terms involving gradients 
of the Higgs field~\cite{tHooft:1974kcl}.}.
One could choose to work in unitary gauge and then
the usual Maxwellian expression for the electromagnetic field holds. 
The advantage of {\it not} working in unitary gauge is that (i) we can treat
$\Phi$ as a dynamical field and it is more straightforward to solve the equations
of motion without going to unitary gauge, and (ii) there are monopole-like
configurations in the electroweak model~\cite{Nambu:1977ag,Achucarro:1999it} 
that make it cumbersome to go to
unitary gauge during the phase transition. These electroweak monopoles 
are singularities in the vector field ${\hat n}^a$. An electroweak monopole is not topological
because it is connected to an antimonopole by a string made of $Z$ magnetic
field. Note that $Z_\mu$ is orthogonal to $A_\mu$ and is defined by,
\be
Z_\mu = \cos\theta_w {\hat n}^a W^a_\mu - \sin\theta_w Y_\mu ,
\label{Zdefn}
\ee
The $Z-$string that confines electroweak monopoles is unrelated to the
electromagnetic magnetic field that emanates from the monopole and
antimonopole at the ends of the string. The configuration is reminiscent
of electrically charged quarks at the end of a gluonic string in QCD.

\subsubsection{Generation of magnetic fields: physical arguments}
\label{Bgeneration}

There are several ways to physically understand the generation of magnetic fields
during the electroweak phase transition and obtain qualitative results with
rough estimates~\cite{Vachaspati:1991nm}. More
quantitative results can be obtained from numerical simulations as described in 
Sec.~\ref{numerical}.

At the electroweak phase transition, the Higgs field $\Phi$ gets a VEV but the orientation 
of the VEV is undetermined. 
The only constraint is $\Phi^\dag \Phi = \eta^2$. Since $\Phi$ has four real degrees
of freedom, the constraint restricts the VEV of $\Phi$ to live on a three sphere 
that can be parametrized as,
\be
\Phi = \eta
\begin{pmatrix}
\sin \alpha\, e^{i\beta}\\
\cos \alpha \, e^{i\gamma}
\end{pmatrix}
\label{3sphereangles}
\ee
where $\alpha \in [0,\pi/2]$, $\beta \in [0,2\pi]$, $\gamma \in [0,2\pi]$ are the (Hopf) 
angular coordinates on the three-sphere. 

Just as in 
the Kibble argument used in the formation of topological defects~\cite{Kibble:1976sj}, the 
Higgs VEV will be different in different spatial domains and $\partial_\mu \Phi \ne 0$ 
in general.
Then the last term in \eqref{Amunu2} will in general not vanish and there is no reason 
to expect the Maxwell field strength, $\partial_\mu A_\nu-\partial_\nu A_\mu$,
to compensate the terms arising from the Higgs field. In fact, because of electroweak 
magnetic monopoles, there are field configurations where compensation is impossible
because the divergence of the Maxwell field strength vanishes while that of the scalar Higgs
does not. Therefore we expect non-zero magnetic fields after the phase transition.

The connection with topological defects can be taken further~\cite{Vachaspati:1994xc}.
We have already discussed that the divergence of magnetic fields in electroweak theory
need not vanish and that there are magnetic monopole configurations in the model. 
For example, a monopole can have the asymptotic field configuration in \eqref{3sphereangles} 
with $\alpha =\theta/2$, $\beta=\phi$ and $\gamma=0$, where $\theta$, $\phi$ are spherical 
angles, and have $\Phi =0$ at their centers~\cite{Nambu:1977ag}.
During the phase transition we can expect that magnetic monopole
configurations will be created but that the monopoles and anti-monopoles
will quickly annihilate since they are confined by Z-strings that will pull them
together. However, a monopole-antimonopole pair (also called a ``dumbbell''~\cite{Nambu:1977ag}) 
has a magnetic dipole field and the annihilation of the dumbbell will release the magnetic fields into 
the ambient plasma. Thus the post-phase transition plasma should be magnetized.

To get an estimate of $E_M(k)$ following from the electroweak phase transition, we will
first estimate $B_{V,\lambda}$ defined in \eqref{BVlambda}. We start by evaluating the volume 
averaged magnetic field of \eqref{BV} using the expression for the magnetic field generated
at the EWPT due to Higgs gradients as in \eqref{Amunu2}. Then,
\ba
(\bfB_V)_i &\sim& 
 - i \frac{2\sin\theta_w}{g\eta^2} \frac{1}{V} \int_V d^3x \,  \epsilon_{ijk} \partial_j\Phi^\dag \partial_k\Phi 
 \nn\\
&\sim &
 - i \frac{2\sin\theta_w}{g\eta^2} \frac{1}{V} \int_{\partial V} dS^{j} \,  \epsilon_{ijk} \Phi^\dag \partial_k\Phi 
 \label{surfaceint}
\ea
where we have performed an integration by parts. 
The final surface integral can be written in terms of the angles on the three-sphere appearing
in \eqref{3sphereangles} under the assumption that $|\Phi |\sim \eta$ on the surface of
integration. The Hopf angles are random variables with probability distributions such
that every point on the three-sphere is equally probable. Since the volume element in Hopf
coordinates is $(1/2) d(\cos(2\alpha)) \, d\beta \, d\gamma$, the variables $\cos(2\alpha )$, $\beta$ 
and $\gamma$ are uniformly distributed over their ranges and take on different values in 
different spatial domains. 
The final surface integral can now be estimated
by assuming that the Hopf angles are approximately constant in domains of size $\xi$. Then
the random elements of the surface integral add up as a random walk and the root-mean-squared
value of the integral goes as the square root of the number of domains on the surface,
\itie~proportional to $\lambda/\xi$ where $\lambda \sim V^{1/3}$ is the characteristic size 
of the volume $V$. Therefore 
\be
B_{V,\lambda} \propto \frac{\lambda}{V} \propto \frac{1}{\lambda^2}  \propto k^2
\label{Best}
\ee
and from \eqref{BVlambda},
\be
E_M(k) \sim \frac{1}{k} B^2_{V,\lambda} \sim k^3.
\label{EMest}
\ee
Note that this estimate using the surface integral does not assume anything
about $|\Phi |$ within the {\it volume} of integration and therefore includes the
possibility of magnetic monopoles within the volume during the EWPT. A direct 
estimate of the volume integral in \eqref{surfaceint} along similar lines would need 
to assume the absence of magnetic monopoles since these necessarily have 
$|\Phi | =0$ at their locations.

Two comments about the estimate in \eqref{EMest} are necessary. 
First, in contrast to our estimate here, the estimate in~\cite{Vachaspati:1991nm} used a
different definition of the ``average magnetic field'' that has been criticized in 
Refs.~\cite{Enqvist:1993np,Hindmarsh:1997tj}. The estimate given here is more
useful because it derives the $k$ dependence of the power spectrum $E_M(k)$ in an unambiguous 
way. The crucial point is that the magnetic field is given by surface (not volume) fluctuations
(see \eqref{surfaceint}).
Second, Ref.~\cite{Durrer:2003ja} argues that causality considerations imply that
fields should have vanishing correlation functions beyond a certain length scale,
and in that case if $E_M(k) \propto k^n$ then necessarily $n \ge 4$, in contradiction 
with the $E_M \propto k^3$ derived above. 
It is certainly true that if correlations in physical space cutoff sharply, the power
spectrum must fall off fast enough for small $k$.
However it is questionable if the premise of the argument 
applies to our situation.
There are several physical systems that have non-vanishing field correlators on arbitrarily 
large distance scales. 
For example, for a free scalar field $\phi$ at temperature 
$T$, the field two-point correlator at large separations is (\iteg~Sec.~3.1 in~\cite{Laine_2016}),
\be
\la \phi (\bfx+\bfr) \phi (\bfx) \ra_T \sim \frac{T}{4\pi |\bfr |} e^{-m |{\bfr}| }, \ \ 
|\bfr | \gg T^{-1}
\ee
where $m$ is the mass of $\phi$. 
The correlator is non-vanishing for all $|\bfr|$; yet this does not signal a violation of 
causality\footnote{Perhaps the ``causality violation'' arises 
in the very act of setting up a large thermal system but this is a basic assumption in cosmology 
where the whole universe is taken to be (essentially) at the same temperature.}. 

A similar calculation can be done for the correlation function of magnetic fields in a thermal
bath of photons with the result, 
\be
\la B_i (\bfx+\bfr) B_j (\bfx) \ra_T = 
- \frac{(rf')'}{r} \delta_{ij} + r \left ( \frac{f'}{r} \right )' {\hat r}_i{\hat r}_j
\ee
where primes denote derivatives with respect to $r$ and
\be
f(r) \equiv  \frac{T}{4\pi r} \coth(\pi T r) 
\ee
The magnetic field correlator clearly does not cutoff sharply beyond some distance and instead 
falls off as $1/r^3$ at large separations. 
To find the power spectrum, $E_M(k)$, we use
\eqref{bbcorr} and express $b_i(\bfk )$ in terms of $B_i(\bfx)$ to get,
\ba
E_M(k) &=& \frac{k^2}{(2\pi)^2} \int d^3x \, e^{i\bfk\cdot\bfx} \la B_i(\bfx)B_i(0) \ra_T \nn \\
&=& \frac{k^3}{(2\pi)^2} \coth \left ( \frac{k}{2T} \right )
\ea
which also derives directly from the Planck distribution, 
$n_k = 1/(e^{k/T}-1)$, noting that $\coth(k/2T) = 2n_k + 1$.
For $k \to 0$, we get $E_M \to k^2 T/(2\pi)^2$. This result makes sense dimensionally
since $E_M$ has mass dimensions of $M^3$ and on physical grounds we expect it to grow 
with $T$. Since long range field correlations exist in the cosmological
medium prior to the electroweak phase transition, it is not surprising that magnetic field 
generation during the transition can lead to long range correlations of the magnetic field.

Another analysis with a system of uncorrelated magnetic dipoles~\cite{Jedamzik:2010cy}
yields magnetic field correlators that are proportional to $k^2$ for small $k$.
In quantum field 
theory, it is well known that vacuum two-point correlators of scalar fields do not vanish at 
arbitrarily large distances. Causality only restricts the expectation of scalar field {\it commutators} 
to vanish beyond the light cone so that no signal can propagate at speeds faster than the
speed of light~\cite{Peskin:1995ev}.

Another argument for the production of magnetic fields during the electroweak phase
transition follows the lines of particle production during reheating after inflation. Before 
the phase transition the VEV of the Higgs field vanishes and the universe is filled with false 
vacuum energy density equal to $\lambda_\Phi \eta^4 =m_H^4/16\lambda_\Phi$
where $m_H\approx 125~\GeV$ is the mass of
the Higgs boson and $\lambda_\Phi \approx 0.13$ is the Higgs quartic coupling constant. 
During the phase transition, the Higgs rolls down its potential and distributes the false 
vacuum energy into energy in other fields. The standard model
has 2 degrees of freedom (d.o.f) for each of the three $W$ gauge bosons and 2 d.o.f.
for the hypercharge gauge field, plus the 4 d.o.f. for the Higgs field, for a total
of 12 d.o.f. in the bosonic sector. If there were only bosonic fields, by equipartition
we would expect each d.o.f. to be populated by about 8\% of the initial energy.
Since the electromagnetic gauge field carries 2 d.o.f., it should carry about
16\% of the initial energy. In reality, we have to take the fermionic d.o.f. into
account as well. Each fermionic family has 4 fermions with a total of 16 d.o.f.
(with massive neutrinos), and with 3 families, we get 48 fermionic d.o.f. in
addition to the 12 bosonic d.o.f., which with equipartition gives us 1.8\%
energy density in each d.o.f.. Assuming equipartition, the electromagnetic field after 
the phase transition should have $\sim 3.6\%$ of the initial energy density. This 
physical argument leads to the expectation that a few percent 
of the cosmic energy density will be in magnetic fields after the electroweak 
phase transition. We can combine this estimate together with \eqref{Best} to write,
\be
E_M(k) \sim \frac{4\,\rho_{EW,B}}{k_*} \left ( \frac{k}{k_*} \right )^3, \ \ k \le k_*
\label{EMEWPT}
\ee
where $\rho_{EW,B} \sim 0.01\,\rho_{EW}$
and $\rho_{EW}$ is the cosmic energy density at the electroweak epoch.
With \eqref{Blambda} this is equivalent to
\be
B_\lambda (t_{\rm EW}) \sim 2 \sqrt{\rho_{EW,B}} \left ( \frac{k}{k_*} \right )^2, \ \ k \le k_*
\label{BlambdaEWPT}
\ee
As in \eqref{Blambda}, $k=2\pi/\lambda$. 
The spectrum peaks at $k=k_*$ and this is where most of the magnetic energy
resides. The corresponding length scale, $\xi \equiv 2\pi/k_*$, will be
referred to as the ``integral scale'' or the ``coherence scale'' of the magnetic field.
The value
of $k_*$ is not known but simulations (see Sec.~\ref{numerical}) indicate that 
it is quite small, of the same order as the inverse lattice size of the simulations. The
value of $k_*$ changes with time during the phase transition; it may also
depend on the assumed nature of the electroweak phase transition.

While these arguments make a strong case for the generation of magnetic
fields at the electroweak phase transition with a significant fraction of the 
cosmic energy density, a more detailed investigation is necessary for confirmation
and to obtain the magnetic field correlation functions. This is why numerical simulations 
of the electroweak phase transition are important. We describe current numerical results
in Sec.~\ref{numerical}.

A difficulty with the electroweak phase transition scenario is that the coherence scale, $\xi$,
is quite small. If we take $\xi$ to be set by the horizon size at the electroweak epoch, 
$t_{\rm EW} \approx 1~\cm$, and the scale comoves with the Hubble expansion,
at the present epoch the coherence scale is merely
$\sim t_{\rm EW} T_{\rm EW}/T_0 \sim 10^{15}\, \cm$. Magnetic fields on scales
smaller than $\sim 1\,\kpc$ ($\sim 10^{21}\, \cm$) today would be dissipated 
(see Sec.~\ref{dissipation}) and
only the tail of the spectrum on scales larger than $\sim 1\,\kpc$ can survive until
today. 
Assuming for the time being the simplest evolution in which magnetic fields
are frozen in the plasma, on a scale of $1\,\kpc$, we find
\be
B_{\rm 1\, \kpc} \sim \sqrt{\rho_{\gamma,0}} \left ( \frac{10^{15}\, \cm}{1\,\kpc} \right )^2
\sim  10^{-18}\, \rmG
\label{Bestnonh}
\ee
where we have used the present epoch energy density in photons
$\rho_{\gamma,0} \sim T_0^4 \sim ( 10^{-6}\, \rmG )^2$ and $T_0 \sim 10^{-4}\,\eV$
is the present photon temperature.

The outcome can change if the magnetic fields generated at the electroweak
phase transition are helical as discussed in Sec.~\ref{mattergenesis}. Then
the magnetic fields can undergo an ``inverse cascade'' in which power from
small length scales is transferred to larger length scales. We will discuss
the evolution of magnetic fields in more detail in Sec.~\ref{evolution}.
For the time being we note that the inverse cascade of helical fields stretches
the coherence scale by an additional 
factor $\sim (T_{\rm EW}/T_{\rm eq})^{2/3} \sim 10^7$ where $T_{\rm eq} \sim 1\, \eV$
is the temperature at cosmic matter-radiation equality.
This brings the comoving\footnote{At time $t_1$, a physical length scale $l(t_1)$ 
corresponds to a ``comoving'' length scale $l_c = l(t_1) a(t_0)/a(t_1)$ where $t_0$ 
is the present cosmological epoch and $a(t)$ is the scale factor of the universe.
Similarly other comoving quantities correspond to their present epoch values
assuming no dynamics other than the expansion of the universe.}
 coherence scale to $\sim 10\,\kpc$. The comoving magnetic 
field strength with the inverse cascade taken into account is smaller than if there were
no inverse cascade by a factor
$\sim (T_{\rm EW}/T_{\rm eq})^{-1/3} \sim 10^{-4}$ giving
\ba
B_{\rm 10\, kpc , \, helical} &\sim& 
\sqrt{\rho_{EW,B}} \left ( \frac{T_0 }{T_{\rm EW}} \right )^2 
\left ( \frac{T_{\rm eq} }{T_{\rm EW}}   \right )^{1/3} \nn \\
&\sim& 10^{-11}\, \rmG
\ea
This estimate shows that helical magnetic fields generated at the EWPT can be strong
enough to produce observed fields in galaxies and clusters.

\subsubsection{Other mechanisms at the EWPT}
\label{othermechs}

There are other mechanisms at the electroweak phase transition that could also 
generate magnetic fields. In the standard model, the electroweak phase transition
is a smooth crossover, but we know the standard model cannot be the full story
because there has to be a mechanism for generating neutrino masses and at
least one dark matter candidate. Then possibly the electroweak phase transition
is first order and the transition to the broken phase occurs by the growth and
nucleation of bubbles of the true vacuum. In Ref.~\cite{Baym:1995fk} the interaction 
of bubble walls with the ambient plasma is argued to generate an electric dipole layer 
on the bubble wall that can result in an electric current and a small magnetic field.
To amplify this field, the authors assume turbulent flows and eventual equipartition
between the kinetic energy density of the flow and the magnetic energy density.
The final estimate for the magnetic field strength is comparable to 
that in \eqref{Bestnonh}.

Another idea related to the electroweak phase transition is that prior to the phase
transition there are {\it pure gauge} configurations that nonetheless are non-trivial,
{\it i.e.} carry non-zero Chern-Simons number. During the electroweak phase transition, 
the Higgs gets a VEV, the gauge fields become massive, and the pure gauge fields (but 
with non-vanishing Chern-Simons number) become physical and decay into electromagnetic 
magnetic fields with helicity~\cite{Jackiw:1999bd}. A follow-up analysis of this process
showed that indeed magnetic fields can be produced by this mechanism but the
magnetic fields are helical only if the Chern-Simons number changes during
evolution~\cite{Zhang_2017}. The production of magnetic fields in this way is similar
to that discussed in Sec.~\ref{Bgeneration} because it depends on the relative misalignment 
in internal space of gauge fields and the vacuum expectation value of the Higgs fields.

\subsubsection{Generation of magnetic fields: numerical results}
\label{numerical}

Several groups have applied numerical techniques to study the electroweak phase
transition and the properties of the magnetic fields that are 
generated~\cite{DiazGil:2007dy,DiazGil:2008tf,Ng:2010mt,Mou:2017zwe,Zhang:2017plw,Zhang:2019vsb}.
So far the simulations have only considered the bosonic sector of the electroweak
model.
Including fermions is of great interest, and there may be novel chiral effects as we
discuss in Sec.~\ref{ideasforamplifying}, but it is a much more difficult endeavor due 
to the inherently quantum nature of fermions~\cite{Aarts:1998td,Borsanyi:2008eu,Saffin:2011kc}.

\begin{figure}
      \includegraphics[width=0.48\textwidth,angle=0]{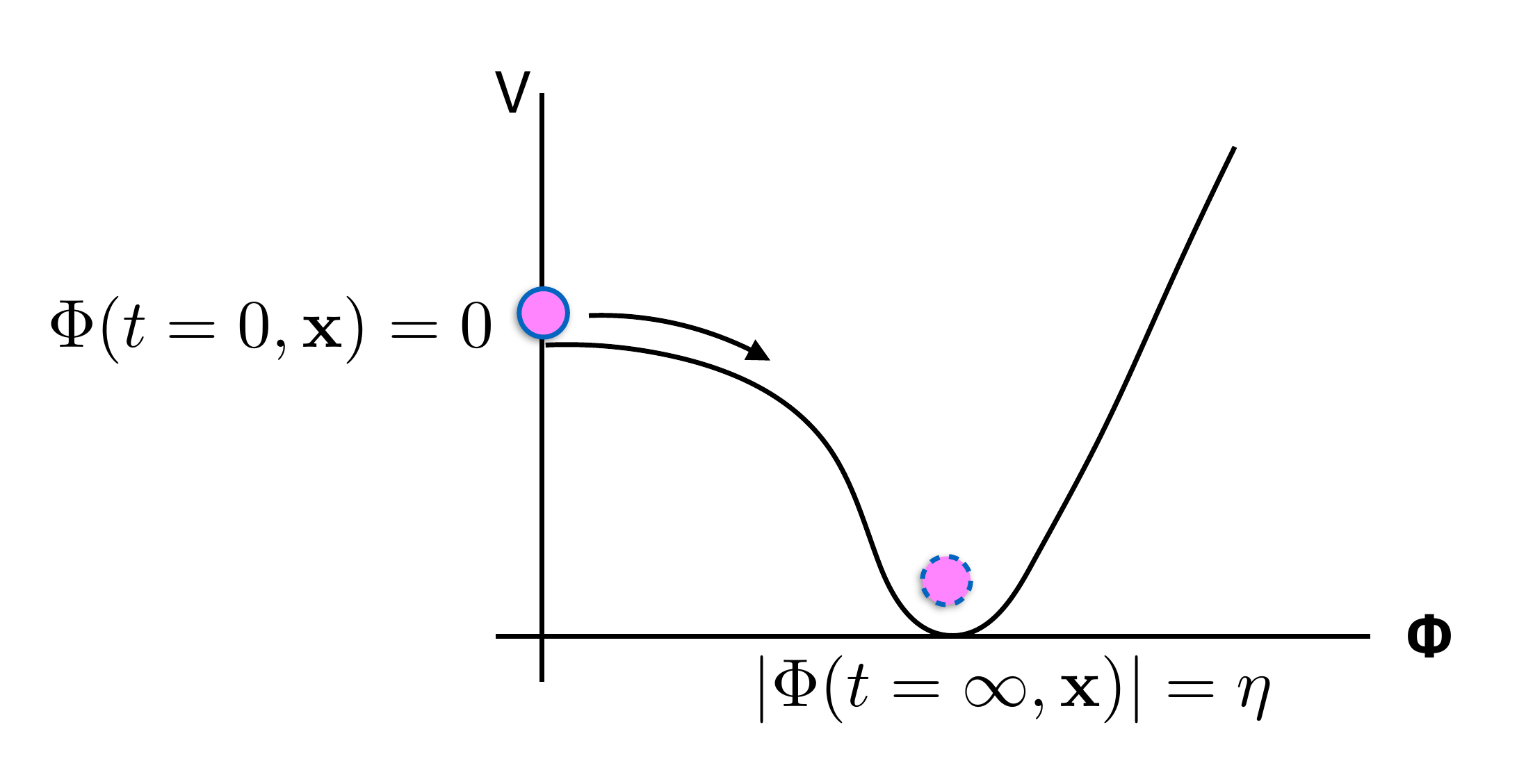}
  \caption{Simplified sketch of the Higgs potential as a function of $|\Phi |$. The Higgs field
  starts at $\Phi=0$ and will roll down to the true vacuum at $|\Phi|=\eta$ which is 
  a three-sphere (angular components of $\Phi$ have been suppressed in the drawing).
  A bubble that triggers the rolling corresponds to a spherical region in space where
  $\Phi \ne 0$ but with no difference in energy as described in Ref.~\cite{Zhang:2019vsb}.}
  \label{higgsPotential}
\end{figure}

\begin{figure}
      \includegraphics[width=0.48\textwidth,angle=0]{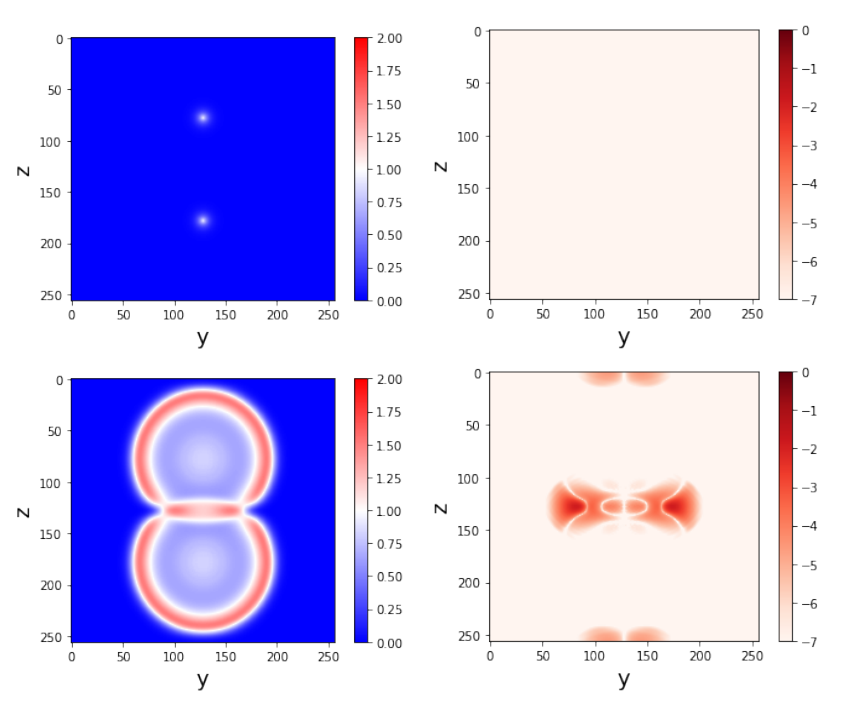}
  \caption{Cross-sectional plot of the magnitude of the Higgs field when two bubbles are 
  nucleated (top left panel) with vanishing magnetic field energy density everywhere
  (top right panel). The bubbles grow and eventually collide (bottom left panel shows the Higgs field
  magnitude) and magnetic fields are produced in the collision region (bottom right panel). Periodic
  boundary conditions are imposed on the lattice, so the bubbles also collide on the
  top and the bottom of the plots.}
  \label{2bubbles}
\end{figure}

The simulations of Ref.~\cite{Zhang:2019vsb} solve the (bosonic) classical electroweak 
equations with initial conditions $\Phi=0$, $W_\mu^a=0=Y_\mu$ and similarly all time
derivatives vanish as well. The Higgs potential is
\be
V(\Phi ) = \lambda_\Phi (\Phi^\dag \Phi - \eta^2)^2
\label{ewpot}
\ee
and the point $\Phi=0$ is a local maximum of the potential (see Fig.~\ref{higgsPotential}).
To trigger the instability where $\Phi$ starts rolling towards
the true vacuum at $|\Phi | = \eta$, ``bubbles'' of vanishingly small energy are
introduced stochastically\footnote{The thorough analysis of 
Ref.~\cite{DiazGil:2008tf} is presented in the context of hybrid inflation but is effectively 
just the electroweak model with a different mechanism for seeding the phase transition.}.
These are not the usual bubbles of a first order phase
transition as the potential is given by \eqref{ewpot} and the point
$\Phi=0$ is unstable, not metastable.
Since $\Phi$ is non-zero inside the bubbles, the system
becomes unstable, $\Phi$ starts rolling towards its true vacuum while also exciting the
other fields, including electromagnetic fields. A crucial point is that $\Phi$ rolls in
different random directions within different bubbles. For example, $\Phi$ could roll
in the $(0,1)^T$ direction in one bubble but $(1+i, 1)^T/\sqrt{3}$ in some other
bubble. When such bubbles collide, (electromagnetic) magnetic fields are produced
as shown in Fig.~\ref{2bubbles}.

\begin{figure}
      \includegraphics[width=0.48\textwidth,angle=0]{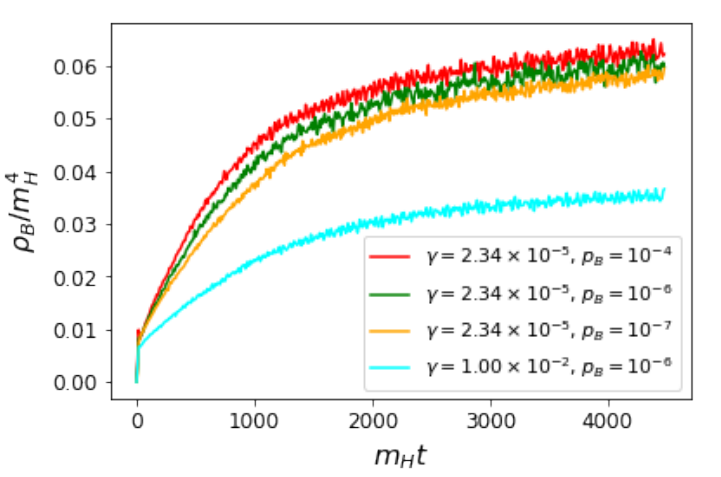}
  \caption{Average energy density in magnetic fields as a function of time for a few
  different choices of damping parameter ($\gamma$) and bubble nucleation
  probability ($p_B$). For details, see~\cite{Zhang:2019vsb}.}
  \label{rhoBvst}
\end{figure}

The next step is to generate a stochastic distribution of bubbles, solve the classical
equations of motion, evaluate the magnetic field and its properties. Fig.~\ref{rhoBvst} 
shows that the average energy density in magnetic fields grows with time and
reaches $\sim 6\%$ of the total energy density by the end of the simulation. (Longer
duration runs were prohibitively expensive.) A point made in Ref.~~\cite{Zhang:2019vsb}
is that most of the magnetic field energy is produced during the time when the 
Higgs field is oscillating around the true vacuum. 

\begin{figure}
      \includegraphics[width=0.5\textwidth,angle=0]{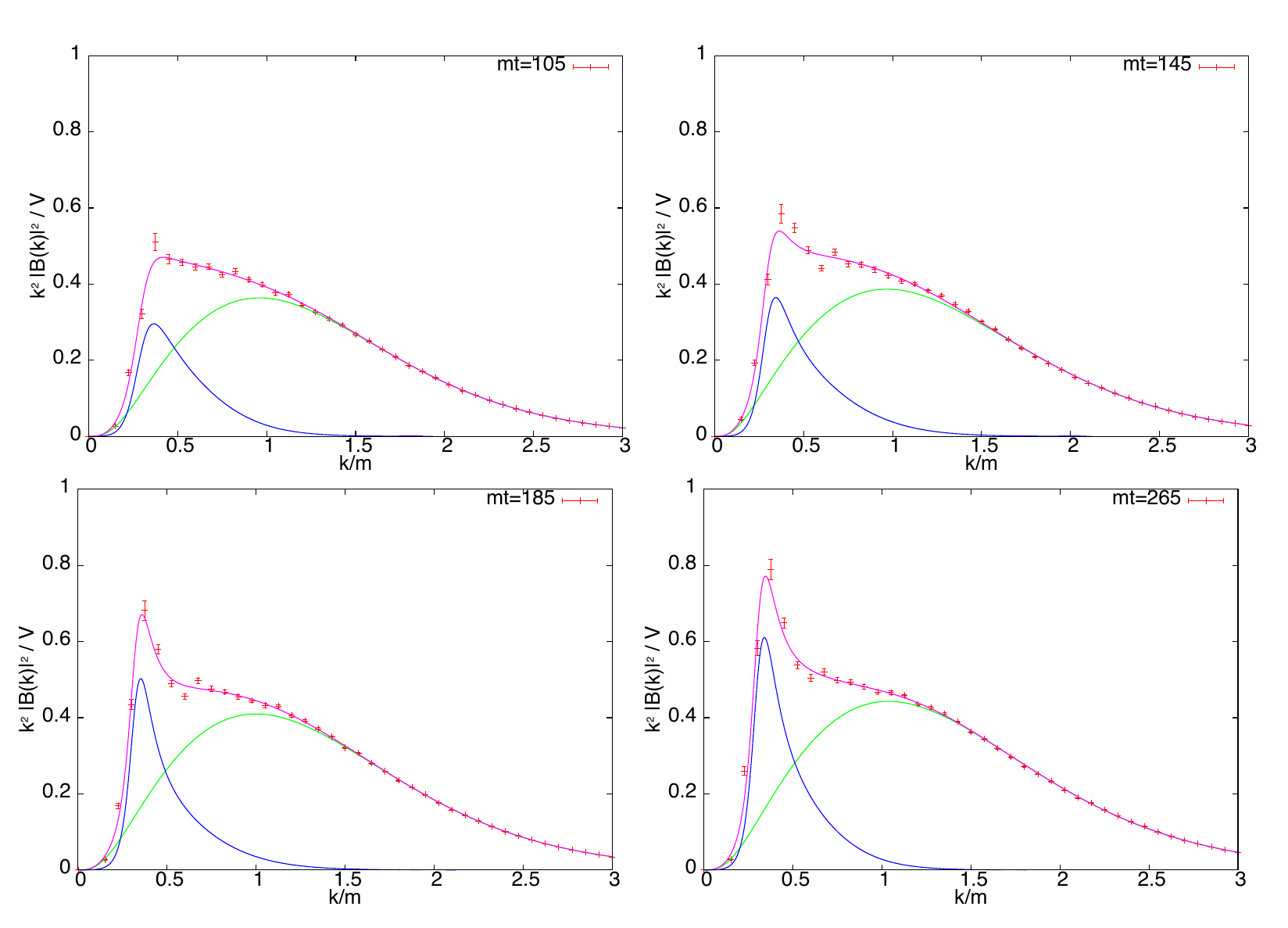}
  \caption{Four snapshots of the power spectrum (pink curve) from the simulations 
  in~\cite{DiazGil:2008tf} show that the peak position ($\approx k_*$ in \eqref{EMEWPT})
evolves to smaller $k$ with time. In this paper
  the authors also decompose the spectrum into a thermal component (green curve) and a
  non-thermal component (blue curve) that is coherent on long length scales. A functional fit 
  that the authors provide gives $E_M(k) \propto k^3$ at small $k$, consistent with the estimate 
  in Eq.~\eqref{EMEWPT}. [Plot from Ref.~\cite{DiazGil:2008tf}.]}
  \label{powerSpectrum}
\end{figure}

Fig.~\ref{powerSpectrum} shows the power spectrum (which is $k^2 |B_k|^2/V$ with the 
conventions of~\cite{DiazGil:2008tf}) at various times in the simulation. 
The peak at very low $k$ implies magnetic field production that is coherent 
on scales comparable to the lattice size and appears in other simulations as well (see
Fig.~13 in Ref.~\cite{Zhang:2019vsb}). According to the discussion 
in~\cite{DiazGil:2008tf}, the peak is generated at high $k$ and then evolves to 
smaller $k$ in what might be termed an ``inverse cascade''. 
It would be of great interest to confirm and understand this phenomenon on analytic 
grounds but it is satisfying that the spectrum has a $k^3$ behavior for small $k$ as
also indicated by Eq.~\eqref{EMEWPT}.

One question is if the energy in electromagnetic fields produced at the EWPT
should be thought of as being in classical electric and magnetic fields or in photons. Though
the quantum particle nature is fundamental, it is legitimate to think of the fields as classical
because we are solving classical equations of motion with the classical expectation 
value of the Higgs field as a source for the gauge fields.
We can also estimate the occupation number of the 
shortest wavelength mode (with $k=k_*$) using Eq.~\eqref{EMEWPT}:
$N_k = (T_{\rm EW}/k_*)^4 \gg 1$ since the (comoving) cut-off length scale on the order of kpc
is vastly larger than the thermal scale on the order of cms. 

Besides the usual limitations of numerical simulations -- small lattices, short run times -- a
strong limitation is that the simulations do not include fermions, a plasma, 
and the expanding cosmological background with a horizon.
These essential elements may affect the ultimate predictions for the properties of the magnetic 
fields produced at the EWPT.

\subsubsection{Magnetic fields and matter genesis}
\label{mattergenesis}

Seemingly different physical phenomena sometimes have an underlying
connection. This is the case for the generation of magnetic field helicity and the creation
of cosmic baryon asymmetry~\cite{Cornwall:1997ms,Vachaspati:2001nb,Garc_a_Bellido_2004}.
Although baryon number is classically conserved in the standard model, a 
quantum anomaly can still violate it. The baryonic current density obeys,
\be
\partial_\mu j^\mu_B = \frac{3}{32\pi^2} \left [ g^2 W^a_{\mu\nu}{\tilde W}^{a\mu\nu} 
                                  - g'^2 Y_{\mu\nu} {\tilde Y}^{\mu\nu} \right ]
\label{anomalyeq}
\ee
where ${\tilde W}^{a\mu\nu} \equiv \epsilon^{\mu\nu\lambda\sigma}W_{\lambda\sigma}/2$.
Now the term on the right-hand side can source baryon number changes.
Eq.~\eqref{anomalyeq} can be integrated over all space to get the change in baryon number
between some initial and final times,
\be
\Delta N_B  = 3 \Delta (CS)
\label{DeltaNB}
\ee
where $CS$ stands for the Chern-Simons number,
\ba
CS &=& \frac{1}{32\pi^2} \int d^3x\, \epsilon_{ijk} \biggl [ \nn \\
&& \hskip -1.3 cm
g^2 \left ( W^{aij} W^{ak} - \frac{g}{3}\epsilon^{abc} W^{ai}W^{bj}W^{ck} \right )
-g'^2 Y^{ij} Y^k \biggr ]
\ea
This formula can also be written in terms of the $W^\pm$,
$Z$ and $A$ (electromagnetism) gauge fields~\cite{Vachaspati:1994ng}. 
Then one gets different
combinations of gauge fields on the right-hand side except for the ${\bf A}\cdot \bfB$
term, \itie~there is no baryon number anomaly due to the electromagnetic 
$\bfE \cdot \bfB$ (or $A{\tilde A}$) in \eqref{anomalyeq}. Then how can changes in baryon 
number be related to the helicity of electromagnetic magnetic fields?

The process of changing the Chern-Simons number can be visualized as the pair
production of a monopole-antimonopole pair (that are connected by a
Z-string in the electroweak model), then the pair is relatively twisted by $2\pi$, and 
allowed to annihilate again~\cite{Vachaspati:1994ng}\footnote{A 
monopole-antimonopole pair with a particular value of the twist is also a solution 
of the electroweak equations and is known as an electroweak 
``sphaleron''~\cite{MantonPhysRevD.28.2019}.}. 
In this process the Chern-Simons number changes by one.
The very fact that monopoles appear in the 
intermediate step means that electromagnetic magnetic fields are present. The
ultimate annihilation of the monopole-antimonopole releases the magnetic field
that also carry the twist. In other words, the magnetic field is helical. Thus
changes in Chern-Simons number are responsible for both baryon number
production and the generation of {\it helical} magnetic 
fields~\cite{Cornwall:1997ms,Vachaspati:2001nb,Garc_a_Bellido_2004}. 
The production of magnetic
fields in Chern-Simons number changing processes has been studied 
in~\cite{Copi:2008he,Chu:2011tx} and during phase transitions 
in~\cite{DiazGil:2007dy,DiazGil:2008tf,Mou:2017zwe,Kharzeev:2019rsy}.

The change in the magnetic helicity density is related to the change in the baryon
number density
\be
\Delta h \sim - \frac{\Delta n_B}{\alpha}
\label{hnB}
\ee
where $\alpha \approx 1/137$ is the fine structure constant and
$h$ is the magnetic helicity density defined in \eqref{hdefn}.
The total helicity is
generally conserved in MHD evolution and hence the helicity density will redshift
with the Hubble expansion as $a^{-3}$ which is identical to the redshifting of
the baryon number density. Since we know the present 
baryon density of the universe, and assuming zero baryon number and 
vanishing magnetic fields initially, \eqref{hnB} gives an estimate for the magnetic 
helicity density produced due to baryon number violation,
\be
h_{\rm b} \sim -10^{-5}\, \cm^{-3}
\label{hbest}
\ee

The discussion above assumes that we start in the {\it broken} phase of the
electroweak symmetry and then a process changes the Chern-Simons number.
During the electroweak phase transition, however, we start in a phase where
the electroweak symmetry is unbroken and electromagnetism isn't even defined.
Then we might expect differences in the numerical estimate of the magnetic helicity
depending on the details of the phase transition.

To test the formation of {\it helical} magnetic fields at the EWPT, the authors 
of Ref.~\cite{Mou:2017zwe} extended the standard model 
by including a CP violating interaction term $|\Phi |^2 W{\tilde W}$ (group and 
Lorentz indices suppressed) in the electroweak Lagrangian and did not
confirm \eqref{hnB}. However, the reason is simple to understand and
might indicate an interesting direction to explore (see Sec.~\ref{ideasforamplifying}).
Once the electroweak symmetry is broken, the $W^a$ gauge fields can be 
re-expressed in terms of the $W^\pm$, $Z$ and electromagnetic $A$ gauge fields. 
Then the extra interaction term explicitly provides an interaction
$|\Phi |^2 A{\tilde A}$ and the dynamics of $\Phi$ can
directly source helicity in the magnetic field. The result
of the simulation is a sum of the helicity due to changes in the 
Chern-Simons number and the helicity due to this direct sourcing and it is
not surprising that the relation between changes of Chern-Simons number 
and magnetic helicity in \eqref{hnB} is not verified. A way around this issue would
be to change the additional interaction term to $|\Phi |^2 (g^2 W{\tilde W} - g'^2 Y{\tilde Y})$ 
as this will eliminate the direct $|\Phi |^2 A {\tilde A}$ coupling. Such a simulation remains 
to be done at this point in time. We will return to this topic in Sec.~\ref{ideasforamplifying}.

Finally let us compare the estimate in \eqref{hnB} with the helicity in inter-galactic
magnetic fields that may be implied by observations. For our numerical estimates 
we will use the most conservative field strength of 
$B_\lambda \sim 10^{-19}\,\rmG$ with $\lambda \sim 1\,\Mpc$~\cite{Finke_2015}.
Since MHD evolution implies that magnetic fields evolve to maximal helicity, we have
with \eqref{realizability}, \eqref{Blambda}, and $k_1=2\pi /\lambda$ where
$\lambda = 3\times 10^{24}\,\cm$,
\be
k_1 H_M(k_1)  =  2 E_M(k_1) = \frac{1}{k_1} B_\lambda^2
\approx 10\, \cm^{-3}
\ee
The helicity density is defined by \eqref{hdefn} and to integrate over all $k$, we need
to assume a spectrum for $H_M$. Let us take $E_M \propto k^{n+1}$ and $H_M \propto k^n$
with electroweak physics suggesting $n=2$ as in \eqref{EMEWPT}.
Then, the observed helicity is at least,
\be
h_{\rm obs} \approx  \frac{k_1 H_M(k_1)}{n+1} \left ( \frac{k_*}{k_1} \right ) ^{n+1}
\sim 10^{3n+3}\, \cm^{-3} \gg h_{\rm b}
\ee
where $k_* \approx 2\pi / (1\, \kpc )$ corresponds to the short distance cutoff due to
dissipation and $h_{\rm b}$ is the helicity resulting from baryogenesis in \eqref{hbest}.
These estimates show that the magnetic helicity produced in association with baryogenesis
is too small to account for the magnetic helicity that must accompany observed magnetic 
fields. This statement comes with the important implication that the standard model of
particle physics may need to be extended to include stronger CP violation if it is to
successfully explain observed inter-galactic magnetic fields, perhaps
along the lines of Sec.~\ref{ideasforamplifying}.

\subsection{Magnetic field generation at QCD and inflationary epochs}
\label{otherideas}

A number of ideas have been proposed for magnetic field generation during the QCD 
epoch~\cite{Quashnock:1988vs,Cheng:1994yr,Sigl:1996dm,Tevzadze:2012kk,Forbes:2000gr,
Miniati_2018}\footnote{Here 
we have to distinguish between ideas for magnetic field
amplification~\cite{Hogan:1983zz} and ideas for generating magnetic fields starting
with no magnetic field. Amplification at the QCD epoch can be important for a pre-existing
magnetic field.}. 
Early ideas~\cite{Quashnock:1988vs,Cheng:1994yr,Sigl:1996dm} were based on the
assumption of a first order QCD phase transition in which bubbles
of the hadronic phase grow within the ambient quark phase. In this process there
is charge separation because the quarks carry a small net positive charge and interact 
with the bubble walls while the leptons carry a small net negative charge and are 
oblivious to the bubbles. It is argued that the dynamics of the bubble walls will
generate turbulence in the charge-separated medium, resulting in electric currents and 
then magnetic fields. An estimate~\cite{Cheng:1994yr} gives $10^{-20}\,\rmG$ on 
a coherence scale $\sim 1\,\kpc$. Scenarios based on a first order QCD phase transition,
however, need to be re-examined as the understanding at present is that the transition
from the quark phase to the hadron phase is a crossover and not a phase
transition ({\it e.g.} see~\cite{Rischke_2004,Fukushima_2010}).
Other ideas to generate magnetic fields invoke ferromagnetic domain walls in 
non-perturbative QCD~\cite{Forbes:2000gr,Tevzadze:2012kk}, and more
recently, axion interactions during the hadronization of quarks 
at the QCD epoch~\cite{Miniati_2018}. 

A number of authors have studied the generation of magnetic fields during inflation by
coupling the inflaton dynamics to that of the electromagnetic field 
(see~\cite{Turner:1987bw,Ratra:1991bn,Adshead:2016iae}
and the reviews~\cite{Widrow:2002ud,Durrer:2013pga,Subramanian:2015lua}).
An advantage is that strong fields can be produced on large coherence scales if certain 
field couplings and interaction strengths are postulated in addition to the usual assumptions
underlying inflation. The disadvantage is that there are few guiding principles that can tell us 
whether the new interactions and coupling strengths are indeed realized. However, if strong,
coherent fields are observed, magnetic fields generated during inflation or its alternatives
may be the only recourse.

\section{MHD evolution of cosmic magnetic fields}
\label{evolution}

The cosmological evolution of magnetic fields is commonly discussed in the MHD
approximation where the displacement current $\partial_t \bfE$ is neglected in 
Maxwell's equations~\cite{Brandenburg:1996fc} (also see Appendix B 
of~\cite{Banerjee:2004df}).
We will restrict our attention to a spatially flat expanding universe for which the
line element can be written as,
\be
ds^2 = a^2(\eta)\, ( d\eta^2 - d\bfx^2 )
\ee
where $a(\eta)$ is the scale factor
and $\eta$ is the conformal time (not to be confused with the VEV of
the Higgs field in earlier sections).
We also define ``comoving'' quantities (denoted by a $c$ subscript),
\be
\bfBc = a^2 \bfB, \  
\bfEc = a^2 \bfE, \ 
\bfv_c=\bfv, \ 
\rho_c = a^4 \rho, \ 
p_c = a^4 p 
\ee
where $\rho$ and $p=\rho/3$ are the energy density and pressure for a
radiation fluid.
Then the equation for the magnetic field is
\be
\partial_\eta \bfBc = \nabla \times (\bfv_c \times \bfBc) + \frac{1}{\sigma_c} \nabla^2 \bfBc
\label{mhdeq}
\ee
where $\eta$ denotes the conformal time, related to the cosmic time $t$ by $dt = a(t) d\eta$,
$\sigma_c =a \sigma$ is the comoving electrical conductivity of the cosmological plasma at 
temperature $T$. The first term on the right-hand side of \eqref{mhdeq} is
called the ``advection'' term. As a plasma element moves, it carries the magnetic field with
it while conserving flux~\cite{Jackson}. 
The second term on the right-hand
side is the ``diffusion'' term since if we ignore the advection term, the MHD equation reduces
to a diffusion equation. The diffusion time scale is set by the electrical conductivity of the
plasma. For large enough electrical conductivity, only the advection
term comes into play and then we say that the magnetic field is ``frozen in'' the plasma.
The evolution can be highly non-trivial even in the frozen in limit because the fluid velocity
field might be turbulent and the magnetic field would get highly tangled, even as the magnetic 
field backreacts on the flow.

The plasma Navier-Stokes equation
for a radiation-baryon fluid of comoving energy density $\rho_c$ and 
comoving pressure $p_c$ in the MHD approximation is~\cite{Brandenburg:1996fc},
\be
 \partial_\eta \bfS =  - (\nabla \cdot \bfv_c ) \bfS
-  (\bfv_c \cdot \nabla )\bfS  - \nabla p_c 
+ (\nabla \times \bfB_c ) \times \bfB_c
\label{navstokes}
\ee
where $\bfS \equiv  (\rho_c + p_c) \gamma^2 \bfv_c$ and 
$\gamma = 1/\sqrt{1-\bfv_c^2}$ is the Lorentz boost factor.
The last term on the right-hand side describes the Lorentz force
on a plasma element due to the magnetic field because $\nabla\times\bfB_c=\bfJ_c$. 
We have ignored the viscosity term and other non-ideal terms but these can be
included.

The continuity equation takes the form,
\ba
\partial_\eta S^0 + \nabla \cdot (S^0 \bfv_c) 
+ \frac{a'}{a} \left ( 4 p_c - \frac{S^0}{\gamma^2} \right ) - \partial_\eta p_c  && \nn \\
&& \hskip -3 cm
= - ( \nabla \times \bfB_c ) \cdot \bfE_c
\label{continuity}
\ea
where $S^0 \equiv (\rho_c+p_c)\gamma^2$ and $a'=da/d\eta$. With $p_c=\rho_c/3$ for 
a radiation fluid, and assuming non-relativistic flows ($\gamma\approx 1$), this equation 
simplifies to
\be
\partial_\eta \rho_c + \frac{4}{3} \nabla \cdot (\rho_c \bfv_c) = -( \nabla \times \bfB_c ) \cdot \bfE_c
\ee
By combining Ohm's law in the rest frame of the fluid with Ampere's law (in the MHD 
approximation) , we can write
\be
\bfE_c = \frac{1}{\sigma_c}\bfJ_c - \bfv_c \times \bfB_c 
= \frac{1}{\sigma_c}(\nabla\times\bfB_c ) - \bfv_c \times \bfB_c .
\label{ohm}
\ee

The electrical conductivity of the cosmological plasma has been estimated
in~\cite{Turner:1987bw,Ahonen:1996nq,Baym:1997gq,Arnold:2000dr}. The
result is
\be
\sigma \sim \frac{T}{\alpha \ln(1/\alpha)}, 
\ee
where the logarithmic correction can be found in~\cite{Baym:1997gq,Arnold:2000dr}.
This calculation assumes that Rutherford scattering dominates and determines the
mean free path. 
Cosmological events such as $e^+e^-$ annihilation at $T\approx 0.1\,\MeV$, 
lead to a sudden drop in the electron number density and then Thomson,
not Rutherford, scattering
determines the mean scattering time. Since $n_e/n_\gamma \approx 10^{-10}$
after annihilation, the electrical conductivity becomes~\cite{Turner:1987bw}
\be
\sigma \sim 10^{-10} \frac{m_e}{e^2}, \ \ T \lesssim 0.1\, \MeV
\ee
However, Rutherford scattering cross section is proportional to $1/v^4 \sim (m_e/T)^2$
while Thomson scattering is independent of $v$. As the universe cools, at 
$T\sim 1\, \eV$, Rutherford scattering dominates again, but now it is
close to the epoch of recombination when the free electron density will once again
drop dramatically by a factor $\sim 10^{-5}$ to that of residual ionization.

In spite of the dramatic drops in electrical conductivity, the magnetic field 
diffusion time scale is much larger than the Hubble time on the length scales 
of interest to us ($\gtrsim 1\,\kpc$ today). The diffusion length scale, $l_{\rm diff}$, 
can be obtained by comparing the left-hand side of \eqref{mhdeq} to the diffusion term 
on the right-hand side,
\be
l_{\rm diff}(t) \sim \sqrt{\frac{t}{\sigma}} = t \, \sqrt{ \frac{\alpha | \ln (\alpha) | }{ tT}}
\ee
(Note that $l_{\rm diff}$ is the physical, not comoving, length scale.)
For example, $l_{\rm diff}(t_0) \sim 10^{12}\,\cm$ where $t_0$ denotes the present
epoch and $l_{\rm diff}(t_{\rm EW}) \sim 10^{-6}\,\cm$
which is much smaller than the electroweak horizon of $\sim 1\,\cm$.
Diffusion is unimportant on length scales larger than $l_{\rm diff}$, while it
must be taken into account in the MHD evolution on length scales smaller than 
$l_{\rm diff}$.

Diffusion isn't the only dissipative mechanism for the magnetic field. Magnetized plasmas
have several excitation modes, called ``magnetosonic'' modes, that can propagate 
and dissipate on various time-scales. 
A tangled magnetic field will excite these modes during evolution and then the modes
will dissipate, with the dissipation time-scale depending on the type of mode.
Without going into details, the dissipation length scale for these modes is 
in the $\pc$ to $\kpc$ range. Thus the power spectrum of cosmological magnetic
fields at the present epoch is expected to be cut off for 
$k \gtrsim 1-10^3\, \kpc^{-1}$~\cite{Jedamzik:1996wp}.

An important result of MHD is that magnetic helicity is conserved during evolution.
This may be proved quite simply in the $\sigma \to \infty$ limit
\ba
\partial_t \int d^3x\, \bfA\cdot \bfB &=&  -2 \int d^3x\, \bfE \cdot \bfB
\nn \\  &=& 
 -\frac{2} {\sigma} \int d^3x\, \bfJ \cdot \bfB \to 0, \ {\rm as}\ \sigma \to \infty \nn
\ea
using Ohm's law as in \eqref{ohm}.
However, the more remarkable feature of
magnetic helicity is that it is conserved in a variety of applications {\it even for
finite conductivity}. Indeed, MHD experts view magnetic helicity conservation
as more robust than energy conservation~\cite{1998pfp..book.....C}. 

The evolution of cosmological magnetic fields has been studied using a
combination of analytical and numerical 
techniques~\cite{Dimopoulos:1996nq,Subramanian:1997gi,Brandenburg:1996sa,Son:1998my,
Christensson:2000sp,Christensson:2002xu,Banerjee:2003xk,Campanelli:2004wm,
Banerjee:2004df,Campanelli:2007tc,Jedamzik:2010cy,Kahniashvili:2010gp,Kahniashvili_2012,
Tevzadze:2012kk,
Kahniashvili:2012uj,Saveliev:2012ea,Campanelli:2013iaa,Saveliev:2013uva,Brandenburg:2014mwa,
Campanelli:2015ypt,Brandenburg:2016odr,
Kahniashvili:2016bkp,Brandenburg:2017rcb,Brandenburg:2017rnt,Brandenburg:2017neh,Reppin:2017uud,
Achikanath_Chirakkara_2021}.
Here we sketch some of the main results, ignoring the finer points and focusing on
magnetic field spectra that fall off on large length scales, $E_M \propto k^n$ with $n > 0$, 
and peak on some small length scale.

\subsection{Sketch of magnetic field evolution\footnote{In the remainder of this section, 
unless otherwise stated, all quantities refer to their comoving values.}}
\label{sketchevolution}

\

{\it Initial conditions}: Stochastic, statistically isotropic magnetic fields are described by two
spectra, $E_M(k,t)$ and $H_M(k,t)$. Any mechanism for generating magnetic fields
should provide these functions at the initial time. Additionally, the velocity field is
also described by two spectra: the power spectrum and the kinetic helicity spectrum.
Even though the set of initial conditions that have been studied is not exhaustive, 
the results give a fairly good idea of what to expect from MHD evolution. 
Ref.~\cite{Banerjee:2004df} also discusses magnetic field evolution through 
cosmological events such as $e^+e^-$ annihilation, recombination, dissipation, etc.
while Ref.~\cite{Brandenburg:2017rnt} considers non-vanishing kinetic helicity as an
initial condition.

\begin{figure}
      \includegraphics[width=0.45\textwidth,angle=0]{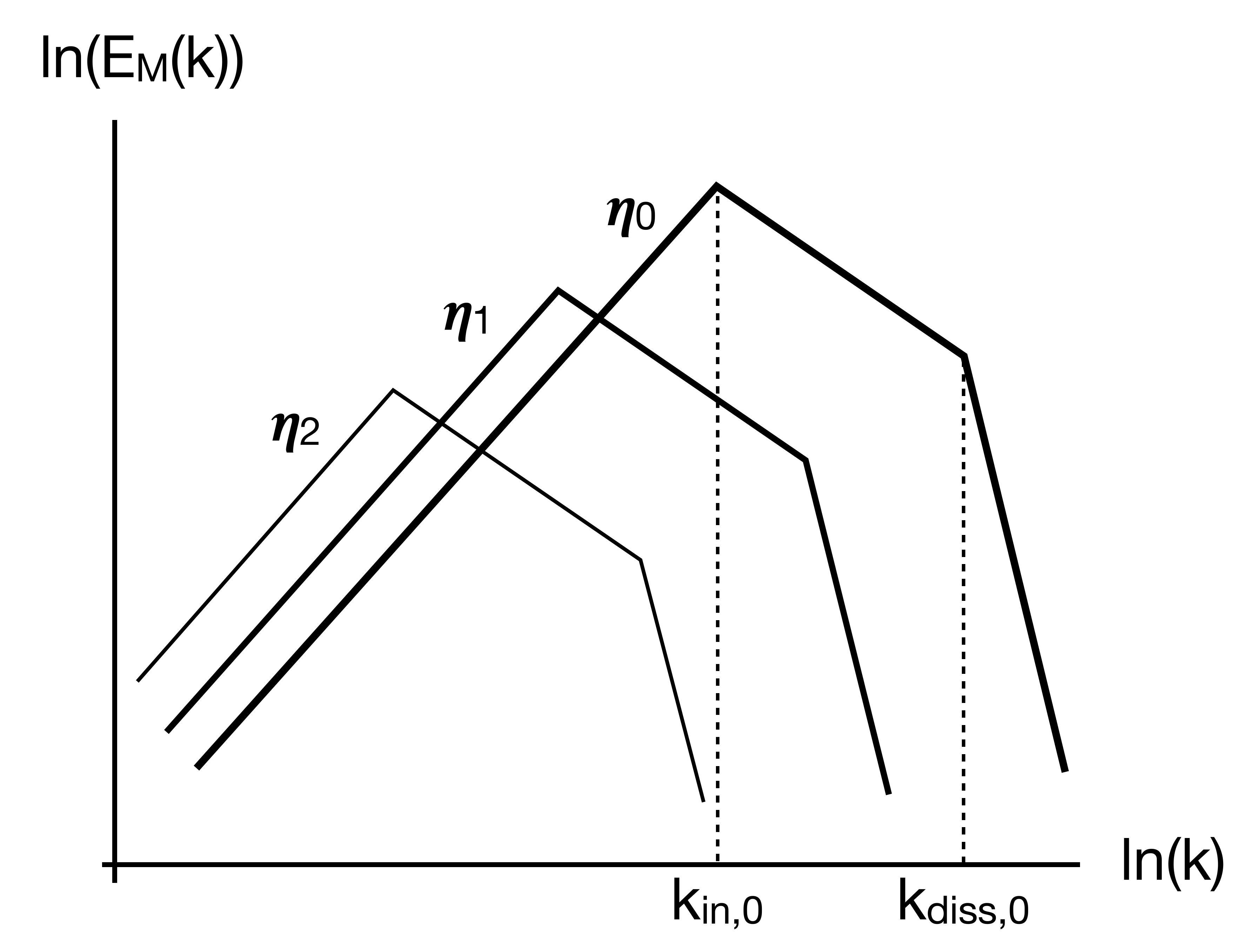}
  \caption{
The evolution of the comoving power spectrum of non-helical magnetic fields shown at three 
(conformal) times $\eta_0 < \eta_1 < \eta_2$. The peak of the spectrum dissipates and also 
moves to smaller (comoving) $k$ in proportion to $\eta^{-1/2}$. As a result the power on large 
length scales increases in what is termed a ``non-helical inverse cascade''. The spectrum 
evolves to $E_M \propto k^4$ at small $k$ for a broad range of initial conditions in which 
$E_M \propto k^p$ with $p \ge 4$ at the initial time.
}
\label{nonhelicalspectrum}
\end{figure}

{\it Non-helical evolution:} 
First consider the case of non-helical magnetic fields, $H_M(k)=0$, and vanishing 
velocity field at the initial time. The magnetic field will induce fluid flow and turbulence,
and the full non-linear system of equations \eqref{mhdeq}, \eqref{navstokes} and 
\eqref{continuity} has to be solved. This setup is called ``magnetically dominated
decaying MHD turbulence'' or ``freely decaying MHD turbulence''.
The detailed evolution of the magnetic power spectrum
depends on the initial conditions. However, the general features of the evolution can
be sketched as in Fig.~\ref{nonhelicalspectrum}.
The peak of the spectrum at the inertial scale, $k_{\rm in}$,
moves to smaller $k$ for two reasons: first, the power on small length scales 
dissipates more efficiently, and second, there is a weak inverse cascade that actually 
shifts the peak to smaller $k$ \cite{Olesen:1996ts,Brandenburg:2014mwa}. These
factors lead to 
\be
k_{\rm in}(\eta) \propto \eta^{-1/2}, \ \ {\rm nonhelical\ case}.
\label{kinnonhelical}
\ee
The $k$-dependence of the spectrum for $k < k_{\rm in}(\eta)$ depends on the initial
conditions. With $E_M \propto k^p$ at the initial time for small $k$ where $p \ge 4$, 
the spectrum evolves to the Batchelor spectrum, $E_M \propto k^4$ (see~\cite{Saveliev:2012ea}
and the Supplemental Material to~\cite{Brandenburg:2014mwa}).  For $p = 2$, called
``non-helical white noise spectrum'', the
simulations in~\cite{Brandenburg:2017neh} do not find an inverse cascade and the 
spectrum stays $E_M \propto k^2$ even upon evolution. This suggests that
for the non-helical case and with $p < 4$, the power spectrum will remain 
$E_M \propto k^p$ even with evolution.

The non-helical inverse cascade has been examined critically in~\cite{Reppin:2017uud}.
It is found that the inverse cascade depends on the Prandtl number, defined as
the ratio of the viscosity and magnetic diffusivity, and decreases in efficiency as the 
Prandtl number increases. Ref.~\cite{Reppin:2017uud} suggests that the non-helical
inverse cascade may be completely suppressed  at the Prandtl numbers relevant
to cosmology ($\sim 10^8 (T/\keV)^{-3/2}$).

If initially the velocity field does not vanish, and is perhaps helical, the evolution can be 
more involved. For example, in case of initial kinetic helicity, the magnetic field can
develop helicity that is opposite to the kinetic helicity on large length scales. The
conservation of total helicity then implies compensating magnetic helicity on short 
length scales~\cite{Brandenburg:2017rnt}.

\begin{figure}
      \includegraphics[width=0.45\textwidth,angle=0]{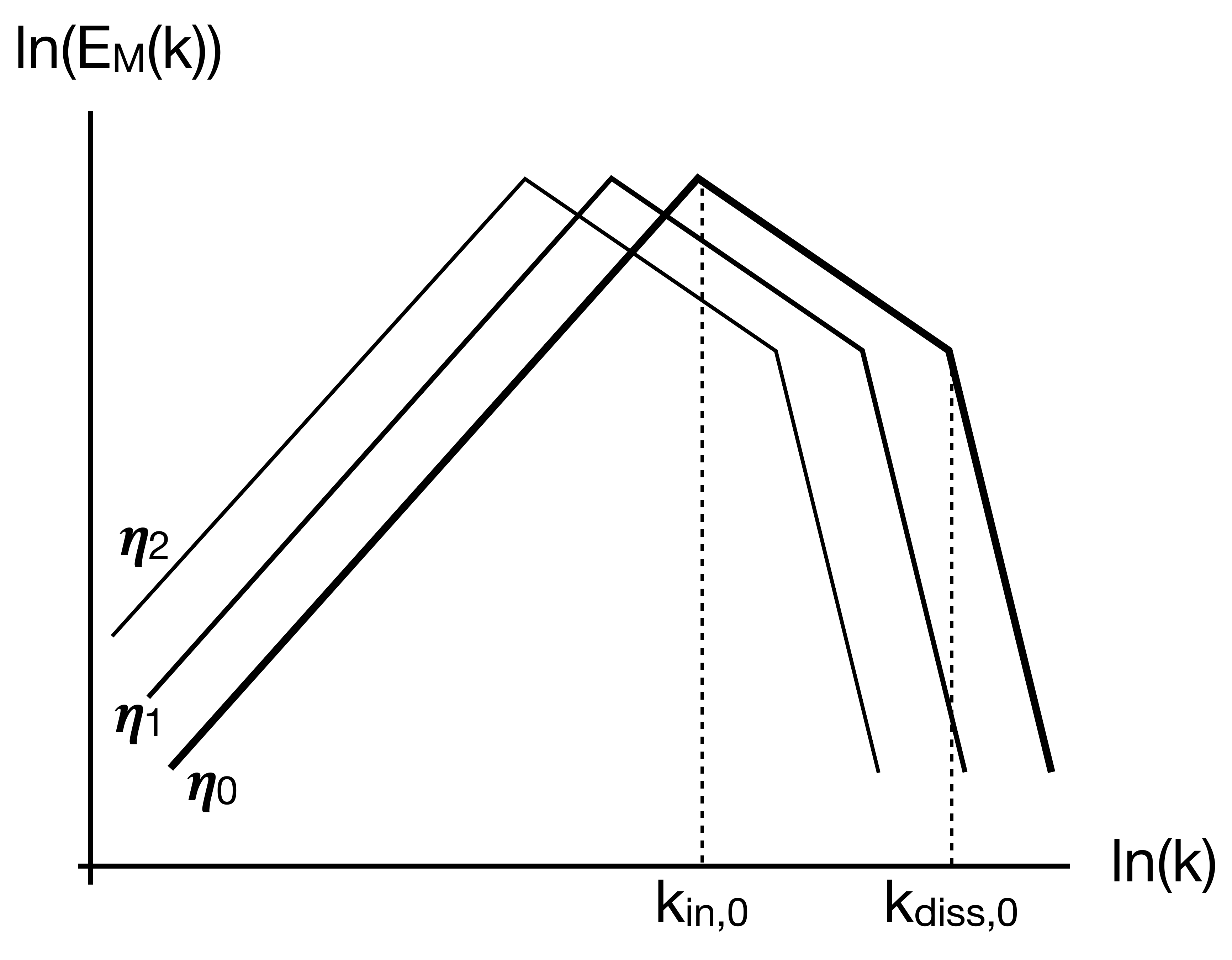}
\caption{
The evolution of the comoving power spectrum of helical magnetic fields at three different
times $\eta_0 < \eta_1<\eta_2$. The peak of the spectrum
moves to smaller comoving $k$ but does not dissipate once the field becomes
maximally helical. As a result the power on large length scales increases
in what is termed an ``inverse cascade''. The spectrum
evolves to $E_M \propto k^4$ at small $k$ for a broad range of
initial conditions in which $E_M \propto k^p$ with $p \ge 4$ at the
initial time.}
\label{helicalspectrum}
\end{figure}

{\it Helical evolution:} 
A general feature is that magnetic fields evolve towards ``maximal helicity''
(see \eqref{realizability}), while total helicity stays conserved.
This means that $H_M(k)$ evolves non-trivially with time while its integral over
all $k$ stays constant. To approach maximal helicity $E_M(k)$ evolves so that
\be
E_M(k,\eta) \to \frac{k}{2} |H_M(k,\eta)|
\ee
Once the field achieves maximal helicity, it stays maximally helical, while 
power (and helicity) is transferred to larger length scales in 
an inverse cascade as depicted in Fig.~\ref{helicalspectrum} where the
whole spectrum simply shifts to the left. 
This feature can be seen using the conservation of magnetic helicity.
We take the form of the time-dependent power spectrum to be
\be
E_M(k,\eta) = {\cal E}(\eta) \left ( \frac{k}{\kin(\eta)} \right )^n, \ \ 0 \le k \le \kin (\eta)
\ee
where $n > 0$ is left unspecified.
(The power for $k > \kin$ can be included but we ignore it here.)
Assuming that the helicity is the same sign for all $k$, maximal helicity implies
\be
H_M(k,\eta) = \frac{2}{k} E_M(k,\eta) 
= \frac{2}{k} {\cal E}(\eta) \left ( \frac{k}{\kin(\eta)} \right )^n .
\ee
The total helicity is conserved,
\be
{\rm constant} = \int_0^{\kin} dk\, H_M(k,\eta) = \frac{2}{n} {\cal E}(\eta)
\ee
Therefore ${\cal E}$ is a constant and the peak of the power spectrum at $k=\kin(\eta)$ 
does not dissipate.
The energy density defined in \eqref{Benergydensity} decays with time as $\kin(\eta)$  
which has been determined by simulations~\cite{Christensson:2000sp,Kahniashvili:2012uj},
\be
k_{\rm in} (\eta) \sim 
k_{\rm in}(\eta_0) \left ( \frac{\eta_0}{\eta} \right )^{2/3}, \ \ {\rm helical\ case}
\label{kintc}
\ee
where $\eta_0$ is the initial conformal time. The $2/3$ exponent in the helical case
should be contrasted with the $1/2$ exponent in the non-helical case given in
\eqref{kinnonhelical}.

As in the non-helical case, the evolution leads to $E_M \propto k^4$
if initially $E_M \propto k^p$ with $p \ge 4$. In simulations with 
magnetic fields that are initially helical with white noise spectrum ($p=2$), 
the spectrum evolved to $E_M \propto k^4$ once the field became
maximally helical, suggesting that the Batchelor 
spectrum is an attractor (see Run G in~\cite{Brandenburg:2017neh}).

\subsection{Dissipation and the integral scale}
\label{dissipation}

The dissipation of perturbations overlaid on a smooth cosmological magnetic field, taking into
account the cosmological plasma and the neutrino and photon fluids, has been studied in Ref.~\cite{Jedamzik:1996wp}. The perturbations can be decomposed into plasma Alfv\'en 
and magnetosonic (fast and slow) modes. These modes will propagate and interact with 
neutrinos prior to neutrino decoupling, and with photons prior to recombination, and
will damp out the perturbations on small wavelengths. The damping length scale
depends on the particular excitation mode, the epoch of interest, and the strength
of the background magnetic field.

The framework of a perturbation on a smooth background magnetic field as 
adopted in~\cite{Jedamzik:1996wp} is not directly relevant to the case of
magnetic fields generated at phase transitions. Instead we are interested in 
power spectra that are peaked on small length scales, {\it e.g.} $E_M \propto k^3$. 
In this case, Banerjee and Jedamzik~\cite{Banerjee:2004df} estimate the 
``integral'' scale of the spectrum -- the length scale on which the spectrum
is peaked -- in the turbulent regime and also in the viscous
regime in which fluid flow experiences a drag due to free streaming particles
such as neutrinos and photons.  
The comprehensive analysis of~\cite{Banerjee:2004df} also
places MHD evolution in a cosmological setting, taking account various cosmological
episodes such as neutrino decoupling, $e^+e^-$ annihilation, Hydrogen recombination, 
and the transition from radiation to matter domination.

The picture developed in Ref.~\cite{Banerjee:2004df} is that there is a direct 
(Kolmogorov) cascade 
of energy for length scales smaller than the integral scale, $L$, and as modes with 
length scale just below $L$ lose their energy, the peak of the spectrum shifts to longer 
length scales, \itie\, the integral scale $L$ grows. Based on a combination of analytical 
and numerical techniques, they argue that the integral scale can be 
estimated by equating the turnover timescale for eddies on the integral scale 
to the Hubble time,
\be
v(L)/L \approx H
\ee
where $v(L)$ is the fluid velocity on the integral scale $L$.
In the turbulent regime, $v(L) \approx v_A$, the Alf\'ven velocity defined by
$v_A=B/\sqrt{4\pi(\rho+p)}$ where $\rho$ and $p$ are the energy density and pressure
of the fluid. In the 
viscous regime, $v(L) \approx R_e v_A$ where $R_e$ is the Reynolds number
evaluated with the Alf\'ven velocity. An important outcome of this analysis is that
the magnetic field strength at the present epoch is related to the
 integral scale today,
\be
B_0 \approx 5 \times 10^{-12}\, \rmG \, \left ( \frac{L_0}{1\,\kpc} \right ) .
\label{BJresult}
\ee
Their final estimate for the integral scale for magnetic fields generated at the EWPT 
depends on the initial magnetic power spectrum and the magnetic helicity, 
and lies in the $0.1\,\pc$ to $10\,\kpc$ range (see Figs.~12-15 in~\cite{Banerjee:2004df}). 
For maximally helical fields with $1\%$ of the total energy density and with $E_M \propto k^2$ 
at the EWPT, they estimate the comoving integral length scale to be $\sim 10\,\kpc$, whereas 
for non-helical fields it is $\sim 1\,\pc$. The magnetic field strength at the integral scale 
can then be evaluated using \eqref{BJresult}.

The evolution of cosmological magnetic fields eventually becomes entangled with the
formation of cosmological structures. This process, together with observational implications, 
has been studied using numerical simulations in Ref.~\cite{Vazza:2017qge}.

\subsection{Speculations on evolution of $E_M \propto k^3$ spectrum}
\label{speculations}

As described in Sec.~\ref{production}, magnetic fields generated at the EWPT are 
expected to have $E_M \propto k^3$ for small $k$ and energy density of a few percent 
of the total energy density. In addition, the magnetic field may be helical though the
fractional helicity is small if it only arises due to baryon number violating processes
(see Sec.~\ref{ideasforamplifying}). The evolution of the integral scale and strength
of magnetic fields with these initial conditions has been studied in Ref.~\cite{Banerjee:2004df} 
with an estimate of $\sim 10\,\pc$ for the integral scale (same as coherence scale), 
and $\sim 10^{-13}\,\rmG$ magnetic field strength at the integral scale. These
estimates depend on the initial magnetic helicity. For initial maximal helicity, the 
integral scale is $\sim 10\,\kpc$ and field strength $\sim 10^{-10}\,\rmG$.

\begin{figure}
      \includegraphics[width=0.44\textwidth,angle=0]{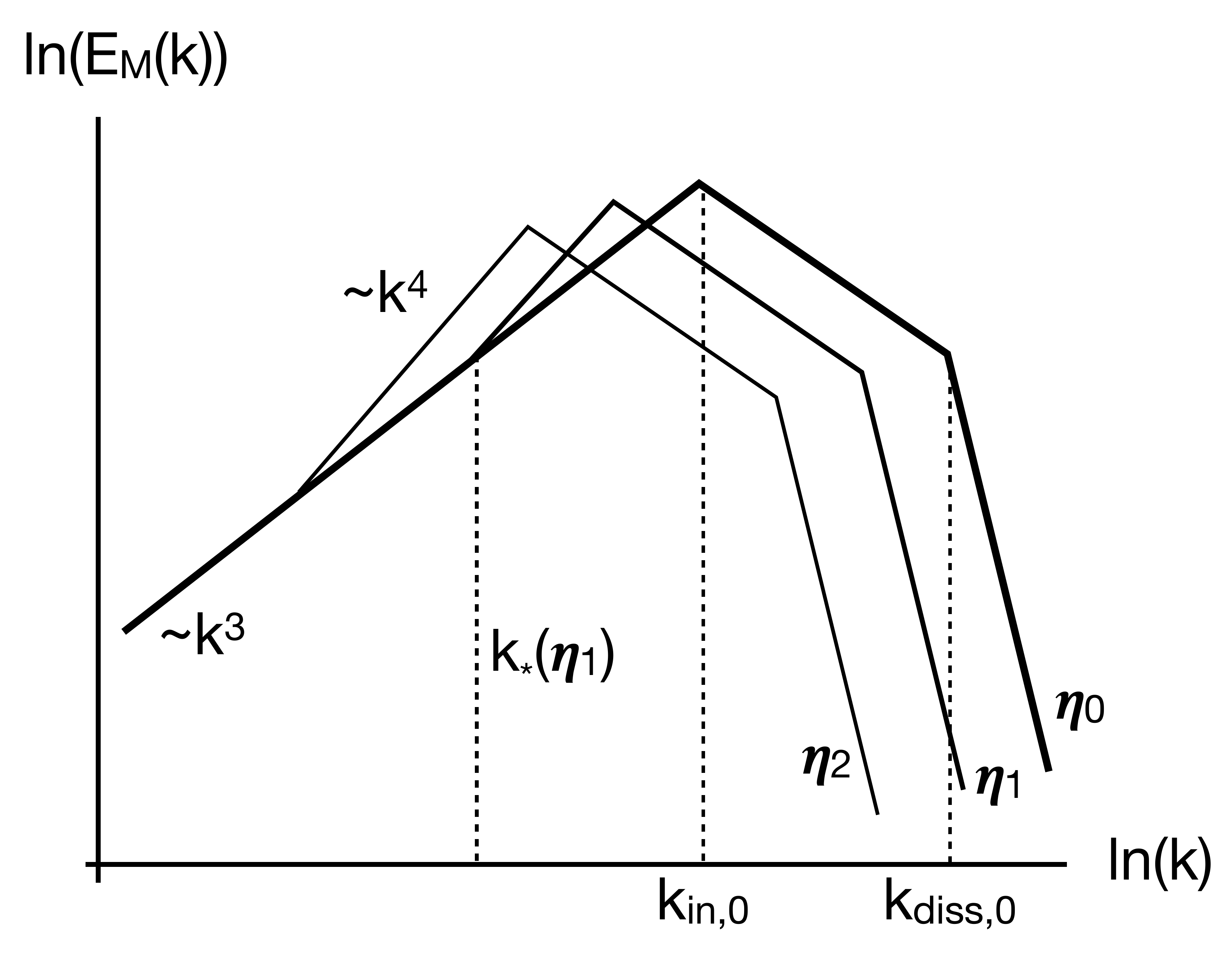}
  \caption{A {\it guess} for the evolution of the comoving $E_M \propto k^3$ spectrum 
  with partial helicity generated at the EWPT 
  at three different times $\eta_0 < \eta_1 < \eta_2$.
  The peak of the spectrum decays
  if there is only partial helicity and moves to the left due to an inverse
  cascade. The helicity in the peak region becomes maximal and the 
  power spectrum in this region develops to $k^4$,
  while the spectrum at the smallest $k$ retains its initial $k^3$ form. 
  The time-dependent $k_*$ denotes the location where the transition from 
  $k^3$ to $k^4$ occurs.}
  \label{spectrumToday}
\end{figure}

The evolution of the power spectrum, $E_M(k,\eta)$, has not been discussed
in Ref.~\cite{Banerjee:2004df}. However, based on the results of~\cite{Brandenburg:2017neh}
we might anticipate some features of the power spectrum.
As an initial condition we assume that $E_M (k,\eta_0) \propto k^3$ up to
$k=k_{\rm in}$. For $k > k_{\rm in}$, the spectrum falls off according to the Kolmogorov
spectrum $\propto k^{-5/3}$, and then a sharper fall off ensues at the dissipation scale
$k_{\rm diss}$ such that modes with $k > k_{\rm diss}$ can release their energy into heat.
At very low $k$, with evolution I expect that the initial spectrum $\propto k^3$ 
will be maintained because these are long length scales and the time scales
will be correspondingly longer. In the very low $k$ region, the evolution
should be as for ``acausal'' spectra in 
Refs.~\cite{Brandenburg:2018ptt,Brandenburg:2020vwp}. 
The integral scale, however, shifts to smaller $k$
as happens in an inverse cascade. For $k$ somewhat
below $k_{\rm in}(\eta)$, we expect the magnetic field to develop towards
maximal helicity and a $k^4$ spectrum~\cite{Brandenburg:2017neh}. For
$k > k_{\rm in}(\eta)$, the Kolmogorov spectrum and the dissipative spectrum
hold. These features are sketched in Fig.~\ref{spectrumToday}. These speculations
imply a new scale that we denote by $k_* (\eta)$ where the $k^3$ spectrum changes 
to a $k^4$ spectrum. With time we expect $k_*$ to move to smaller $k$ to 
maintain the $k^4$ power law for $k_* < k < k_{\rm in}$.
If the magnetic field is near maximally helical at the initial time, it may also
happen that the peak at $k_{\rm in}(\eta)$  actually grows in amplitude as the
integral scale shifts to lower $k$. It would be of great interest to examine this
case in more detail by numerical simulations.

\subsection{Qualitative constraint plot}
\label{constraintplot}

\begin{figure}
      \includegraphics[width=0.48\textwidth,angle=0]{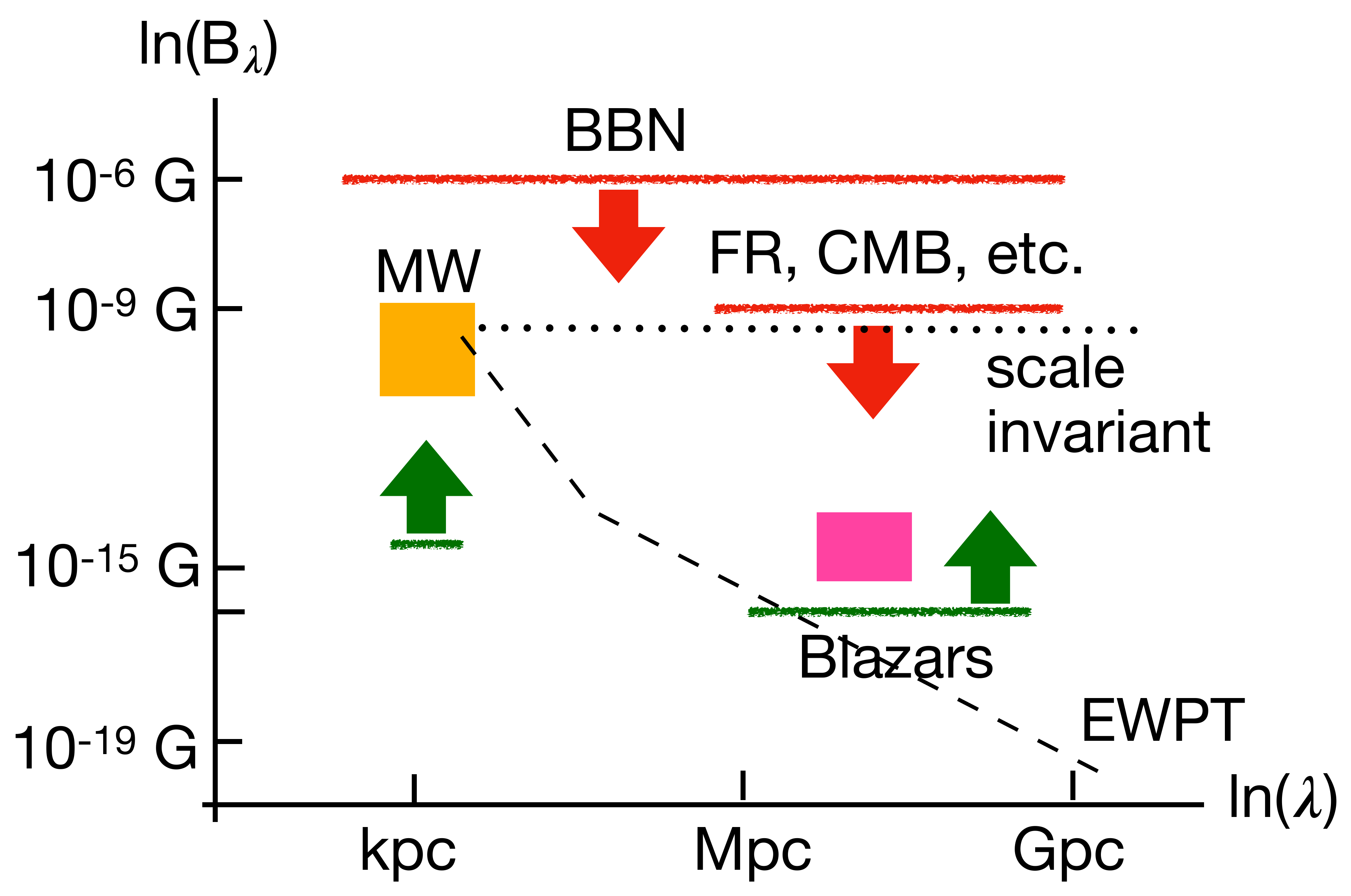}
  \caption{
Cosmological magnetic fields with 
$\lambda \sim \kpc$ and $B_\lambda \sim 10^{-10}\,\rmG$ may directly explain,
\itie~with minimal dynamo amplification, the galactic magnetic field, and may help
resolve the present Hubble tension due to baryon clustering at Hydrogen recombination. 
This region is denoted by the golden rectangle marked by ``MW'' (for Milky Way).
Big bang nucleosynthesis (BBN) constrains
$B_\lambda \lesssim 10^{-6}\,\rmG$ on all scales, while other observations
roughly constrain $B_\lambda \lesssim 10^{-9}\,\rmG$ on Mpc to Gpc scales.
Blazar spectral measurements place a lower bound 
$\sim 10^{-16}\,\rmG$ for $\lambda$ in the Mpc to Gpc range, 
or instead a lower bound of $\sim 10^{-15}\,\rmG$ on kpc scales,
and assumes that plasma instabilities do not play a role.
Magnetic helicity measurements are uncertain but, if
confirmed, would lie in the region of the pink rectangle. If the cosmological
magnetic field is scale invariant, as in some inflationary models, and also passes
through the golden rectangle, it would correspond to the horizontal dotted line and 
is pressured by several constraints. If the cosmological magnetic field has a blue
$k^3$ or $k^4$ spectrum as might be expected from the EWPT, it may correspond 
to a shape like the dashed curve. This constraint plot is meant to be schematic as
there is considerable uncertainty in the numbers and also the caveats mentioned
in the text. For example, $B_\lambda$ may be below the blazar bound shown by the
green line at $\lambda \sim \Mpc$ since small scale (kpc) magnetic fields can instead 
explain the absence of blazar halos.
}
  \label{constraints3}
\end{figure}

Ideally a constraint plot maps out the regions in the $\lambda$-$B_\lambda$ plane
(equivalently the $k$-$E_M(k)$ plane because of \eqref{Blambda})
that are disallowed or else suggested by observations. However, as discussed in 
Sec.~\ref{observations}, to translate an observational constraint to a region of
the $\lambda -B_\lambda$ plane, we need to know the window function and
this is not always available and may not be simple. Instead a constraint may be 
derived by comparing
observations to the results of simulations that are done by assuming a certain
form of $E_M(k)$, \iteg~the magnetic field has fixed field strength and is uniform 
in domains of size 1~Mpc. Hence a constraint plot should be treated with a
certain amount of caution. Nonetheless a constraint plot can be very helpful to 
quickly visualize the overall status of cosmological magnetic fields.
With this disclaimer, we sketch current constraints and other markers
in the $\lambda-B_\lambda$ plane in Fig.~\ref{constraints3}.

\section{Amplifying magnetic helicity}
\label{ideasforamplifying}

We have seen in Sec.~\ref{production} that magnetic fields can be generated during the
EWPT with relatively large energy density and with $k^3$ power spectrum. Without
magnetic helicity though, the evolution discussed in Sec.~\ref{sketchevolution} shows 
that the field strength on large ($\sim\Mpc$) length scales will be very weak.
Magnetic helicity can significantly change these estimates as it modifies the 
evolution by producing
an inverse cascade that pushes the coherence scale to larger length scales and also 
increases the power on these scales. The connection
of magnetic field generation with baryogenesis described in Sec.~\ref{production} shows
that the standard model has all the ingredients necessary for generating magnetic helicity.
However the numbers don't quite work out as discussed in Sec.~\ref{mattergenesis} and
the magnetic helicity produced during baryogenesis falls short of what is needed by many
orders of magnitude. This failure, as well as the failure of the standard model to explain
the cosmic baryon asymmetry, the dark matter problem and neutrino masses, suggests 
that the standard model needs to be extended.

One possibility relevant to magnetic fields is that the electroweak plasma 
contains chiral fermions with net chirality
~\cite{vachaspati2021cosmological}. 
Then the chiral-vortical~\cite{Vilenkin:1979ui} and
chiral-magnetic~\cite{Vilenkin:1980fu} effects may come into 
play, effectively providing an electric current that is proportional
to the magnetic field and this can amplify the magnetic 
helicity~\cite{Joyce:1997uy,tashiroPhysRevD.86.105033,Dvornikov:2013bca,Dvornikov:2016jth,
Rogachevskii:2017uyc,
Brandenburg:2017rcb,Schober:2017cdw}. However, as noted in
Ref.~\cite{Brandenburg:2017rcb} the helicity amplification is limited by the assumed initial 
chirality of the medium and is further limited due to chirality 
flipping~\cite{boyarsky2020evolution,boyarsky2020equilibration}.
Alternately, as we have mentioned in Sec.~\ref{sketchevolution}, 
kinetic helicity can separate magnetic helicity of  opposite signs. Perhaps in chiral scenarios
this can result in amplified magnetic helicity on large length scales that is approximately 
compensated by opposite magnetic helicity on small scales such that the net magnetic helicity 
is not too large.

Yet another possibility is that there are new CP violating interactions 
in particle physics. 
A simple way to introduce additional CP violation in the standard model is to add the interactions,
\ba
L_{\rm CP} &=& \frac{|\Phi|^2}{\eta^2}  \biggl [ \beta_{\rm CS} \left ( 
c_w^2 W_{\mu\nu}^a {\tilde W}^{a\mu\nu} - s_w^2 Y_{\mu\nu} {\tilde Y}^{\mu\nu} \right ) \nn \\
&& \hskip 0.5 cm
+ \beta_A  \left ( 
s_w^2 W_{\mu\nu}^a {\tilde W}^{a\mu\nu} - c_w^2 Y_{\mu\nu} {\tilde Y}^{\mu\nu} \right ) \biggr ]
\label{LCP}
\ea
where $\beta_{\rm CS}$ and $\beta_A$ are coupling strengths, 
$s_w^2 \equiv \sin^2\theta_w \approx 0.23$, $c_w\equiv \cos\theta_w$, 
and $\eta$ is the VEV of $\Phi$.
A reason for the particular grouping of terms in \eqref{LCP} is that the Chern-Simons (CS) combination
is natural for baryogenesis, while the other combination is natural for magnetic helicity
generation. To see this, let us work in the unitary gauge where $\Phi \propto (0,1)^T$, suppress
Lorentz indices, and assume $W^1=0=W^2$. Then we can
rewrite the groupings in terms of $Z_{\mu\nu}$ and $A_{\mu\nu}$ (written as $Z$ and $A$),
\ba
c_w^2 W^a {\tilde W}^a - s_w^2 Y{\tilde Y}&=& c_{2w} Z {\tilde Z} + s_{2w} A{\tilde Z}
\nn \\
s_w^2 W^a {\tilde W}^a - c_w^2 Y{\tilde Y}&=& -c_{2w} A{\tilde A} + s_{2w}A{\tilde Z}
\nn
\ea
where $c_{2w}=\cos(2\theta_w)$ and $s_{2w}=\sin(2\theta_w)$. The Chern-Simons combination
of terms directly couples the Higgs to $Z{\tilde Z}$ but not to $A{\tilde A}$, while the other 
combination provides a direct coupling to the electromagnetic $A{\tilde A}$ but not to $Z{\tilde Z}$.
As the phase transition proceeds and the Higgs rolls down on its potential, the first group of
interactions in $L_{\rm CP}$ will generate changes in the
Chern-Simons number that will lead to baryogenesis via \eqref{DeltaNB} and (anomalous) 
helical magnetic field production as discussed in \eqref{hnB}. The second combination of interactions
in $L_{\rm CP}$ will directly generate helical magnetic fields. In fact the simulations of 
Ref.~\cite{Mou:2017zwe} were done with $\beta_A/\beta_{\rm CS} = - \tan^2\theta_w$ and 
hence included only the $W{\tilde W}$ terms in $L_{\rm CP}$.
The simulations resulted in magnetic helicity production that was uncorrelated with the
baryon number production. This is not surprising since with this choice of coefficients
there is direct magnetic helicity production due to the $\beta_A$ term in \eqref{LCP}
and also weaker helicity production due to the indirect coupling through changes in
the Chern-Simons number. It is of immense interest to investigate these simulations
more thoroughly to determine if there is a sweet spot that can accommodate both the baryon 
number of the universe and the helical magnetic fields indicated by blazar observations. If there 
is such a sweet spot, the observation of cosmological magnetic fields might lead to the
discovery of CP odd operators in fundamental particle interactions, another beautiful
inner space/outer space connection!

The additional couplings in $L_{\rm CP}$ are constrained by collider data as they
provide new channels for the Higgs boson to decay into two photons, and to
a $Z$ boson and a photon. At the moment 
the experimental constraints are roughly $\beta_{\rm CS},\beta_A \lesssim 10^{-2}$
as in Ref.~\cite{Gan_2017}\footnote{In ~\cite{Gan_2017} the authors also list other dimension 
six operators, including another CP violating operator 
$\Phi^\dag \sigma^a \Phi W^a_{\mu\nu}{\tilde Y}^{\mu\nu}$ where $\sigma^a$ are the
Pauli spin matrices, that could be relevant to magnetic fields.}. 
Constraints on $\beta_{\rm CS}$ and $\beta_A$ from the observation 
of magnetic fields and baryon asymmetry would provide helpful guidance to particle
experiments that should ultimately measure these parameters~\cite{Bishara_2014}.

It is difficult to predict the magnetic helicity spectrum that will be induced by the terms in 
\eqref{LCP} without detailed analyses. However it is still useful to make a guess
that can serve as a strawman for when such analyses are carried out. We have already
discussed that the dynamics of the electroweak symmetry breaking leads to significant
magnetic fields. Such fields are not helical. With the $L_{\rm CP}$ extension, additional
magnetic fields are generated due to the $\beta_A$ coupling. Such magnetic fields
are helical. The $\beta_{\rm CS}$ coupling also gives rise to helical magnetic fields
but this contribution is small. So the overall picture is that the $\beta_A$ coupling
leads to helical magnetic fields that are diluted by the non-helical fields produced by
the Higgs dynamics in the un-extended standard model. A simple guess
is that the magnetic helicity spectrum is proportional to the power 
spectrum with the constant of proportionality determined by the coupling constant
$\beta_A$,
\ba
H_M(k) &=& \beta \frac{2}{k} E_M(k) \nn \\
&=& \frac{2\beta \rho_{EW,B}}{k_*^2} \left ( \frac{k}{k_*} \right )^2, \ \ k \le k_*
\label{HMEW}
\ea
where we have used \eqref{EMEWPT} in the second line, and $\beta \sim \pm 10^{-2}$ 
represents the coupling constants and the sign will depend on the sign of $\beta_A$. It 
would be of interest to study the evolution of cosmological magnetic fields with the initial 
conditions given by \eqref{EMEWPT} and \eqref{HMEW}.

\section{Concluding remarks}
\label{conclusions}

Remarkable progress has been made on cosmological magnetic fields in the last
few decades. Observations have placed upper and lower bounds on the magnetic
field strength on large length scales (Sec.~\ref{observations}). The effects of magnetic 
fields on Hydrogen 
recombination and the implications for cosmological observations have begun to 
be investigated. Simulations of the electroweak phase transition show the production
of magnetic fields that decay relatively slowly on large length scales, establishing 
a novel observational probe of particle physics (Sec.~\ref{production}). The MHD 
evolution of cosmological magnetic fields has been mapped out leading to a broad 
brush picture through various cosmological epochs and enabling the prediction of 
observational consequences (Sec.~\ref{evolution}).

As we press forward in the investigation of cosmological magnetic fields, a number
of questions present themselves. The primary observational evidence for magnetic
fields is the absence of blazar halos but this may also be explained if there are
plasma instabilities (Sec.~\ref{plasmainstability}). The direct detection of halos in 
stacked blazar data (Sec.~\ref{halodetection}) 
argues against plasma instabilities but consensus is still needed on claims
of halo detection. Can the parity-odd helicity of the magnetic field help in their
detection (Sec.~\ref{searchforhelicity})?
Magnetic fields at the recombination epoch may relieve the 
Hubble tension (Sec.~\ref{Brecomb}) but how can we confirm this resolution? 
Are there other CMB signatures, perhaps some spectral signatures?
While it is clear that the electroweak phase transition generates magnetic fields,
the survival depends on the helicity of the magnetic field (Sec.~\ref{production}). 
Is this an indication of CP violating physics beyond the standard model? It is exciting that 
magnetic field observations can perhaps motivate detectable signatures
in accelerator experiments (Sec.~\ref{ideasforamplifying}). 

Finally, the magnetic fields generated during the electroweak phase transition
may affect subsequent cosmological events, \iteg\ the QCD phase transition.
Conversely, the magnetic field spectra carry imprints of every cosmological 
episode between now and the time of magnetic field generation. 
If we are able to detect and measure the power spectra of cosmological magnetic 
fields, the ``fine structure'' of the spectra could become a novel probe of the early universe.

\acknowledgements

I am grateful to
Robi Banerjee, Matthew Baumgart,
Gordon Baym, Heling Deng, Ruth Durrer, Francesc Ferrer, Juan Garcia-Bellido, 
Karsten Jedamzik, Tina Kahniashvili, Mikko Laine, Andrew Long, Andrii Neronov, 
Teerthal Patel, Levon Pogosian, Andrey Saveliev, Igor Shovkovy, and George Zahariade
for their comments and feedback.

TV is supported by the U.S. Department of Energy, Office of High Energy Physics, 
under Award DE-SC0019470 at ASU.

\newpage

\bibstyle{aps}
\bibliography{review}

\end{document}